\documentclass[12pt]{article}
\usepackage{latexsym,amssymb,amsfonts,amsmath}
\usepackage{mathrsfs} 
\usepackage[dvips]{graphicx}
\newlength{\extraspace}
\newlength{\extraspaces}
\newcommand{\be}{\begin{equation}
\addtolength{\abovedisplayskip}{\extraspaces}
\addtolength{\belowdisplayskip}{\extraspaces}
\addtolength{\abovedisplayshortskip}{\extraspace}
\addtolength{\belowdisplayshortskip}{\extraspace}}
\newcommand{\ee}{\end{equation}}

\newcommand{\bea}{\begin{eqnarray}
\addtolength{\abovedisplayskip}{\extraspaces}
\addtolength{\belowdisplayskip}{\extraspaces}
\addtolength{\abovedisplayshortskip}{\extraspace}
\addtolength{\belowdisplayshortskip}{\extraspace}}
\newcommand{\eea}{\end{eqnarray}}

\newcommand{\newsection}[1]{
\vspace{15mm}
\pagebreak[3]
\addtocounter{section}{1}
\setcounter{equation}{0}
\setcounter{subsection}{0}
\setcounter{footnote}{0}
\begin{flushleft}
{\Large\bf \thesection. #1}
\end{flushleft}
\nopagebreak
\medskip
\nopagebreak}

\newcommand{\newsubsection}[1]{
\vspace{1cm}
\pagebreak[3]
\addtocounter{subsection}{1}
\noindent{\bf \thesubsection. #1}
\nopagebreak
\vspace{2mm}
\nopagebreak}

\newcommand{\La}{\mathcal{L}}
\newcommand{\OO}{\mathcal{O}}

\newcommand{\sgru}{SU(L)\otimes SU(L)}

\newcommand{\csti}{F^{2}_{\pi}+3F^{2}_{X}}

\newcommand{\demu} {\partial_{\mu}}
\newcommand{\demup} {\partial^{\mu}}
\DeclareMathOperator{\Tr}{Tr}
\newcommand{\fp} {F_{\pi}}
\newcommand{\fn} {F_{\eta'}}
\newcommand{\fx} {F_{X}}
\newcommand{\sq} {\sqrt{2}}
\newcommand{\mn} {m_{\eta}}
\newcommand{\mx} {m_{\eta_X}}
\newcommand{\mnn} {m_{\eta'}}
\newcommand{\mpi} {m_{\pi}}
\newcommand{\dd} {\mathrm{d}}
\newcommand{\mev} {\mathrm{MeV}}
\newcommand{\kev} {\mathrm{keV}}
\newcommand{\gev} {\mathrm{GeV}}

\begin{document}

\addtolength{\baselineskip}{.8mm}

{\thispagestyle{empty}

\noindent \hspace{1cm} \hfill Revised version \hspace{1cm}\\
\mbox{}                \hfill February 2014 \hspace{1cm}\\

\begin{center}
\vspace*{1.0cm}
{\large\bf EFFECTS OF AN EXTRA $U(1)$ AXIAL CONDENSATE ON}\\
{\large\bf THE STRONG DECAYS OF PSEUDOSCALAR MESONS}\\
\vspace*{1.0cm}
{\large Enrico Meggiolaro\footnote{E--mail:
enrico.meggiolaro@df.unipi.it} }\\
\vspace*{0.5cm}{\normalsize
{Dipartimento di Fisica, Universit\`a di Pisa,
and INFN, Sezione di Pisa,\\
Largo Pontecorvo 3,
I--56127 Pisa, Italy.}}\\
\vspace*{2cm}{\large \bf Abstract}
\end{center}

\noindent
We consider a scenario (supported by some lattice results) in which a
$U(1)$--breaking condensate survives across the chiral transition in QCD.
This scenario has important consequences for the pseudoscalar--meson sector,
which can be studied using an effective Lagrangian model.
In particular, generalizing the results obtained in two previous papers,
where the effects on the radiative decays $\eta,\eta' \to \gamma\gamma$
were studied, in this paper we study the effects of the $U(1)$ chiral
condensate on the strong decays of the ``light'' pseudoscalar mesons, i.e.,
$\eta,\eta' \to 3\pi^0$; $\eta,\eta' \to \pi^+ \pi^- \pi^0$;
$\eta' \to \eta \pi^0 \pi^0$; $\eta' \to \eta \pi^+ \pi^-$;
and also on the strong decays of an exotic (``heavy'') $SU(3)$--singlet
pseudoscalar state $\eta_X$, predicted by the model.

\vspace{0.5cm}
}
\newpage

\newsection{Introduction}

\noindent
It is well known that the QCD vacuum has a very complicated structure,
characterized by some non--trivial local (or also non--local) {\it condensates},
whose behaviour as a function of the temperature $T$ also characterizes the
phase structure of the theory.\\
For example, a phase transition which occurs in QCD at a finite temperature
$T_{ch}$ is the restoration of the $SU(L) \otimes SU(L)$ chiral symmetry
(in association with $L=2,3$ massless quarks), which for $T < T_{ch}$
is broken spontaneously by the non--zero value of the so--called
``{\it chiral condensate}'', i.e., $\langle \bar{q} q \rangle \equiv
\sum_{i=1}^{L} \langle \bar{q}_i q_i \rangle$ \cite{chiral-symmetry}.
But QCD with $L$ massless quarks has also (at least at the classical level)
a $U(1)$ axial symmetry \cite{Weinberg1975,tHooft1976}. This symmetry is
broken by an anomaly at the quantum level, which in the
``{\it Witten--Veneziano mechanism}'' \cite{Witten1979a,Veneziano1979}
plays a fundamental role (via the so--called ``{\it topological
susceptibility}'') in explaining the large mass of the $\eta'$ meson.
The role of the $U(1)$ axial symmetry for the finite temperature phase 
structure has been so far not well clarified. One expects that, above a
certain critical temperature $T_{U(1)}$, also the $U(1)$ axial symmetry
will be (effectively) restored but it is still unclear whether $T_{U(1)}$
has or has not something to do with $T_{ch}$.

In this paper we re--consider a scenario (which was originally proposed in
Refs. \cite{EM1994a,EM1994b,EM1994c,EM1995} and elaborated in Refs.
\cite{EM2002,EM2003,EM2004}, and which seems to be supported by some
lattice results on the so--called ``{\it chiral susceptibilities}''
\cite{Bernard-et-al.1997,lattice2000,lattice2010}) in which a new
$U(1)$--breaking condensate survives across the chiral transition at $T_{ch}$,
staying different from zero up to a temperature $T_{U(1)} > T_{ch}$.
$T_{U(1)}$ is, therefore, the temperature at which the $U(1)$ axial symmetry
is (effectively) restored, meaning that, for $T>T_{U(1)}$, there are no
$U(1)$--breaking condensates.
The new $U(1)$ chiral condensate has the form
$C_{U(1)} = \langle {\cal O}_{U(1)} \rangle$,
where, for a theory with $L$ light quark flavours, ${\cal O}_{U(1)}$ is a
$2L$--fermion local operator that has the chiral transformation properties of
\cite{tHooft1976,KM1970,Kunihiro2009}:\footnote{Throughout this paper we use
the following notations for the left--handed and right--handed quark fields:
$q_{L,R} \equiv \frac{1}{2} (1 \pm \gamma_5) q$,
with $\gamma_5 \equiv -i\gamma^0\gamma^1\gamma^2\gamma^3$.}
\be
{\cal O}_{U(1)} \sim \displaystyle{{\det_{st}}(\bar{q}_{sR}q_{tL})
+ {\det_{st}}(\bar{q}_{sL}q_{tR}) },
\label{Ou1}
\ee
where $s,t = 1, \ldots ,L$ are flavour indices. The colour indices [not
explicitly indicated in Eq. (\ref{Ou1})] are arranged in such a way that:
{\it i)} ${\cal O}_{U(1)}$ is a colour singlet, and {\it ii)}
$C_{U(1)} = \langle {\cal O}_{U(1)} \rangle$ is a {\it genuine} $2L$--fermion
condensate, i.e., it has no {\it disconnected} part proportional to some
power of the quark--antiquark chiral condensate $\langle \bar{q} q \rangle$:
the explicit form of the condensate for the cases $L=2$ and $L=3$ is discussed
in detail in Appendix A (see also Refs. \cite{EM1994c,EM1995,EM2002}).

This scenario has important consequences for the pseudoscalar--meson sector.
The low--energy dynamics of the pseudoscalar mesons, including the effects due
to the anomaly, the $q\bar{q}$ chiral condensate and the new $U(1)$ chiral
condensate, can be described, in the limit of large number $N$ of colours,
and expanding to the first order in the light quark masses, by an effective
Lagrangian written in terms of the topological charge density $Q$, the mesonic
field $U_{ij} \sim \bar{q}_{jR} q_{iL}$ (up to a multiplicative constant) and
the new field variable $X \sim {\det} \left( \bar{q}_{sR} q_{tL} \right)$
(up to a multiplicative constant), associated with the new $U(1)$ condensate
\cite{EM1994a,EM1994b,EM1994c,EM2002}:
\bea
\lefteqn{
\La (U,U^\dagger ,X,X^\dagger ,Q)
= \frac{1}{2}\Tr(\partial_\mu U\partial^\mu U^\dagger )
+ \frac{1}{2}\partial_\mu X\partial^\mu X^\dagger } \nonumber \\
& & -V(U,U^\dagger ,X,X^\dagger)
+ \frac{i}{2}\omega_1 Q \Tr(\ln U - \ln U^\dagger) \nonumber \\
& & + \frac{i}{2}(1-\omega_1)Q(\ln X-\ln X^\dagger) + \frac{1}{2A}Q^2,
\label{lagrangian}
\eea
where the potential term $V(U,U^{\dagger},X,X^{\dagger})$ has the form:
\bea
\lefteqn{
V(U,U^\dagger ,X,X^\dagger )
= \frac{\lambda_{\pi}^2}{4} \Tr[(U^\dagger U
-\rho_\pi {\bf I})^2] +
\frac{\lambda_X^2}{4} (X^\dagger X-\rho_X )^2 } \nonumber \\
& & -\frac{B_m}{2\sqrt{2}}\Tr(MU+M^\dagger U^\dagger)
-\frac{c_1}{2\sqrt{2}}[\det(U)X^\dagger + \det(U^\dagger )X].
\label{potential}
\eea
$M={\rm diag}(m_1,\ldots,m_L)$ is the quark mass matrix
and $A$ is the topological susceptibility in the pure Yang--Mills (YM) theory.
(This Lagrangian generalizes the one originally proposed in Refs.
\cite{DV1980,Witten1980,RST1980,KO1980,NA1981}, which included only the
effects due to the anomaly and the $q\bar{q}$ chiral condensate.)
All the parameters appearing in the Lagrangian must be considered as 
functions of the physical temperature $T$. In particular, the parameters 
$\rho_{\pi}$ and $\rho_X$ determine the expectation values $\langle U \rangle$
and $\langle X \rangle$ and so they are responsible respectively for the
behaviour of the theory across the $SU(L) \otimes SU(L)$ and the $U(1)$ chiral
phase transitions, as follows:
\bea
\rho_\pi|_{T<T_{ch}} &\equiv& \frac{1}{2} F_\pi^2 > 0, ~~~
\rho_\pi|_{T>T_{ch}} < 0, \nonumber \\
\rho_X|_{T<T_{U(1)}} &\equiv& \frac{1}{2} F_X^2 > 0, ~~~
\rho_X|_{T>T_{U(1)}} < 0.
\label{table}
\eea
The parameter $F_\pi$ is the well--known pion decay constant, while the
parameter $F_X$ is related to the new $U(1)$ axial condensate.
Indeed, from Eq. (\ref{table}),
$\rho_X = \frac{1}{2} F_X^2 > 0$ for $T<T_{U(1)}$, and therefore, from Eq.
(\ref{potential}), $\langle X \rangle = F_X/\sqrt{2} \ne 0$. Remembering that
$X \sim {\det} \left( \bar{q}_{sR} q_{tL} \right)$, up to a multiplicative
constant, we find that $F_X$ is proportional to the new $2L$--fermion
condensate $C_{U(1)} = \langle {\cal O}_{U(1)} \rangle$ introduced above.
In the same way, the pion decay constant $F_{\pi}$, which controls the breaking
of the $\sgru$ symmetry, is related to the $q\bar{q}$ chiral condensate 
by a simple and well--known proportionality relation (see Refs.
\cite{EM1994a,EM2002} and references therein):
$\langle \bar{q}_i q_i \rangle_{T<T_{ch}} \simeq - \frac{1}{2}B_m F_\pi$.
(Moreover, in the simple case of $L$ light quarks with the same mass $m$,
$m^2_{NS} = m B_m/F_\pi$ is the squared mass of the non--singlet pseudoscalar
mesons and one gets the well--known Gell-Mann--Oakes--Renner relation:
$m^2_{NS} F^2_\pi \simeq -2m \langle \bar{q}_i q_i \rangle_{T<T_{ch}}$.)
It is not possible to find, in a simple way, the analogous relation between
$F_X$ and the new condensate $C_{U(1)} = \langle {\cal O}_{U(1)} \rangle$.

However, as was shown in two previous papers \cite{EM2003,EM2004}, information
on the quantity $F_X$ (i.e., on the new $U(1)$ chiral condensate, to which it
is related) can be derived, in the realistic case of $L=3$ light quarks with
non--zero masses $m_u$, $m_d$ and $m_s$, from the study of the radiative
decays of the pseudoscalar mesons $\eta$ and $\eta'$ into two photons.
A first comparison of the results with the experimental data
has been performed and it is encouraging, pointing towards some
evidence for a non--zero $U(1)$ axial condensate.\\
The following decay rates are derived \cite{EM2003,EM2004}:
\bea
\label{gammaeta}
\Gamma(\eta\to\gamma\gamma) &=&
\frac{\alpha^{2}m_{\eta}^{3}}{192\pi^{3}F_{\pi}^{2}}\left(\cos\tilde{\varphi}+
\frac{2\sq F_{\pi}}{F_{\eta'}}\sin\tilde{\varphi}\right)^{2}, \\
\label{gammaeta'}
\Gamma(\eta'\to\gamma\gamma) &=&
\frac{\alpha^{2}m_{\eta'}^{3}}{192\pi^{3}F_{\pi}^{2}}
\left(\frac{2\sq F_{\pi}}{F_{\eta'}}\cos\tilde{\varphi}
-\sin\tilde{\varphi}\right)^{2},
\eea
where $\alpha=e^{2}/4\pi \simeq 1/137$ is the fine--structure constant.
Here $F_{\eta'}$ is defined as follows:
\be
F_{\eta'} \equiv \sqrt{\csti},
\label{F-eta'}
\ee
and can be identified with the $\eta'$ decay constant in the chiral limit of
zero quark masses. Moreover, $\tilde{\varphi}$ is a mixing angle, which can be
related to the masses of the quarks $m_u$, $m_d$, $m_s$, and therefore to the
masses of the octet mesons, by the following relation:
\be
\tan{\tilde{\varphi}}= \frac{\sqrt{2}}{9A}B \fp \fn (m_s - \tilde{m}) =
\frac{\fp \fn}{6 \sqrt{2}A} (m_{\eta}^2 - m_{\pi}^2),
\label{phitilde}
\ee
where:
$m^{2}_{\pi}=2B\tilde{m}$ and $m_{\eta}^{2}=\frac{2}{3}B(\tilde{m}+2m_s)$,
with: $B \equiv \frac{B_{m}}{2F_{\pi}}$ and
$\tilde{m} \equiv \frac{m_u+m_d}{2}$.\\
If one puts $F_X=0$, i.e., if one neglects the
new $U(1)$ chiral condensate, the expressions written above reduce to the
corresponding ones derived in Ref. \cite{DNPV1981} using
an effective Lagrangian which includes only the usual $q\bar{q}$ chiral
condensate.
Using the experimental values for the various quantities which appear in Eqs.
(\ref{gammaeta}) and (\ref{gammaeta'}),
one can extract the following values for the quantity $F_X$ and for the
mixing angle $\tilde\varphi$:\footnote{Indeed, the
original values reported in Refs. \cite{EM2003,EM2004} were:
$F_X = 27(9)$ MeV and $\tilde\varphi = 16(3)^\circ$. The values reported
in Eq. (\ref{results}) (which are, anyhow, consistent with the original
values within the errors) have been obtained using the updated experimental
values of the \emph{Particle Data Group} \cite{PDG2010} (in particular:
$\Gamma_{\rm exp}(\eta\to\gamma\gamma) = 0.51(3)$ keV and
$\Gamma_{\rm exp}(\eta'\to\gamma\gamma) = 4.31(36)$ keV; moreover we use:
$F_\pi = 92.2(4)$ MeV, $m_\pi \simeq 134.98$ MeV, $m_\eta \simeq 547.85$ MeV,
$m_{\eta'} \simeq 957.78$ MeV).}
\be
F_X = 24(7) ~{\rm MeV},~~~ \tilde\varphi = 17(2)^\circ,
\label{results}
\ee
and these values are perfectly consistent with the relation (\ref{phitilde})
for the mixing angle, if one uses for the pure--YM topological susceptibility
the estimate $A=(180\pm5~\rm{MeV})^{4}$, obtained from lattice simulations
\cite{lattice}.

In the section 3 of this paper, continuing the work started in Refs.
\cite{EM2003,EM2004}, we shall study the effects of the $U(1)$ chiral
condensate on the strong decays of the ``light'' pseudoscalar mesons, i.e.,
$\eta,\eta' \to 3\pi^0$; $\eta,\eta' \to \pi^+ \pi^- \pi^0$;
$\eta' \to \eta \pi^0 \pi^0$; $\eta' \to \eta \pi^+ \pi^-$;
and also on the strong decays of an exotic (``heavy'') $SU(3)$--singlet
pseudoscalar state $\eta_X$, predicted by the model:
$\eta_X \to 3\pi^0$; $\eta_X \to \pi^+ \pi^- \pi^0$;
$\eta_X \to \eta \pi^0 \pi^0$; $\eta_X \to \eta \pi^+ \pi^-$;
$\eta_X \to \eta' \pi^0 \pi^0$; $\eta_X \to \eta' \pi^+ \pi^-$;
$\eta_X \to 3\eta, 3 \eta', \eta\eta\eta', \eta\eta'\eta'$.
In particular, in the case of the exotic particle $\eta_X$, we shall
find some relations between its mass and its decay widths, which in principle
might be useful to identify a possible candidate for this particle.

For the benefit of the reader, we shall start, in section 2, by resuming the
main results, obtained in the original papers \cite{EM1994a,EM1994c,EM2002},
concerning the mass spectrum of the Chiral Effective Lagrangian
(\ref{lagrangian})--(\ref{potential}), for temperatures $T < T_{ch}$:
in this paper we shall consider the case $T=0$ only.

\newpage

\newsection{Mass spectrum and new parameters of the Chiral Effective Lagrangian}

\noindent
Let us consider the Lagrangian \eqref{lagrangian}, where the field variable
$Q(x)$ has been integrated out:
\bea
\label{newlagr}
\lefteqn{
\La (U,U^\dagger ,X,X^\dagger) =
\frac{1}{2} \Tr (\demu U \demup U^{\dagger}) +
\frac{1}{2} \demu X \demup X^{\dagger} - V(U,U^\dagger ,X,X^\dagger) }
\nonumber \\
& & + \frac{1}{8}A \left[ w_1 \Tr (\ln U - \ln U^{\dagger}) +
(1-w_1) (\ln X - \ln X^{\dagger}) \right]^2 .
\eea

\newsubsection{Mass spectrum at $T=0$ for a generic $L$ (in the chiral limit)}

\noindent
At $T=0$ both $SU(L) \otimes SU(L)$ and $U(1)_A$ symmetries are broken.\\
Following Ref. \cite{DV1980}, we can eliminate the redundant (having much
larger masses) scalar fields of the {\it linear} $\sigma$--type model by
taking the limit $\lambda_{\pi}^2 \rightarrow \infty$ and
$\lambda_X^2 \rightarrow \infty$. In this limit the potential term gives the
following constraints:
\be
U^{\dagger} U = \frac{1}{2} \fp^2 \cdot {\bf{I}} \ ;
\quad X^{\dagger} X = \frac{1}{2} \fx^2 .
\label{vincoli}
\ee
We are thus left with a {\it non--linear} chiral effective model, in which
the field $U$ has the form:
\be
U= \sqrt{\frac{1}{2}} \fp \exp \left \{ \frac{i \sqrt{2}}{\fp}\Phi \right \},~~~
\Phi= \sum_{a=1}^{L^2-1}{ \pi_a \tau_a + \frac{S_{\pi}}{\sqrt{L}} {\bf{I}} } ,
\label{Uexp}
\ee
where $\tau_a \ (a=1, \cdots, L^2-1)$ are the generators of $SU(L)$
($\Tr(\tau_a)=0$) in the fundamental representation, with normalization
$\Tr(\tau_a \tau_b)=\delta_{ab}$, and $\pi_a \ (a=1, \cdots, L^2-1)$ are the
non--singlet meson fields, while $S_{\pi}$ is the {\it usual} quark--antiquark
$SU(L)$ singlet field:
\be
S_{\pi} \sim i \sum_{i=1}^L{\left(\bar{q}_{iL} q_{iR} -
\bar{q}_{iR}q_{iL} \right)}.
\ee
And similarly the field $X$ has the form:
\be
X= \sqrt{\frac{1}{2}} \fx \exp \left \{ \frac{i \sqrt{2}}{\fx} S_X \right \}
\label{Xexp}
\ee
where $S_X$ is an {\it exotic} singlet field, with the following quark content:
\be
S_X \sim i [\det_{st}(\bar{q}_{sL} q_{tR}) - \det_{st}(\bar{q}_{sR} q_{tL})].
\ee
Substituting Eqs. \eqref{Uexp} and \eqref{Xexp} into Eq. \eqref{newlagr}
and taking only the quadratic part of the Lagrangian, we obtain:
\bea
\La_2 &=& \frac{1}{2} \partial_{\mu}\pi_a \partial^{\mu} \pi_a + \frac{1}{2} \partial_{\mu} S_{\pi} \partial^{\mu}S_{\pi} + \frac{1}{2} \partial_{\mu} S_{X} \partial^{\mu}S_{X} - \frac{1}{2} \left(\sum_{il}{\mu_i^2 \tau_{il}^a \tau_{li}^b} \right) \pi_a \pi_b \nonumber \\
&-& \frac{1}{2} \left( \frac{2}{\sqrt{L}}\sum_i{\mu_i^2\tau_{ii}^a} \right)
\pi_a S_{\pi}- \frac{1}{2L} \sum_i{\mu_i^2 S_{\pi}^2} \nonumber \\
&-& \frac{1}{2}c \left(\frac{\sqrt{2L}}{\fp} S_{\pi}- \frac{\sqrt{2}}{\fx}S_X \right)^2 - \frac{1}{2}A \left[ \frac{\sqrt{2L}}{\fp}\omega_1 S_{\pi} + \frac{\sqrt{2}}{\fx}(1-\omega_1)S_X \right]^2 ,
\label{lagrquad}
\eea
where:
\be
c \equiv \frac{c_1}{\sqrt{2}} \left( \frac{\fx}{\sqrt{2}}\right)
\left(\frac{\fp}{\sqrt{2}}\right)^L ,~~~
\mu_i^2 \equiv \frac{B_m}{\fp} m_i .
\label{c-mu}
\ee
In the {\it chiral limit}, $\sup m_i \rightarrow 0$, Eq. \eqref{lagrquad}
reduces to:
\bea
\La_2 &=& \frac{1}{2} \partial_{\mu}\pi_a \partial^{\mu} \pi_a + \frac{1}{2} \partial_{\mu} S_{\pi} \partial^{\mu}S_{\pi} + \frac{1}{2} \partial_{\mu} S_{X} \partial^{\mu}S_{X} \nonumber \\
&-& \frac{1}{2}c \left(\frac{\sqrt{2L}}{\fp} S_{\pi}- \frac{\sqrt{2}}{\fx}S_X \right)^2 
-\frac{1}{2}A \left[ \frac{\sqrt{2L}}{\fp}\omega_1 S_{\pi} + \frac{\sqrt{2}}{\fx}(1-\omega_1)S_X \right]^2.
\label{lagrquadchirale}
\eea
In this case the $L^2-1$ non--singlet fields are massless: they are the
Goldstone bosons coming from the breaking of the $SU(L) \otimes SU(L)$ symmetry
down to $SU(L)_V$. Instead, the two singlet fields $S_{\pi}$ and $S_X$ are
mixed with the following squared mass matrix:
\be
\begin{pmatrix}
\frac{2L(A \omega_1^2+c)}{\fp^2} & \frac{2 \sqrt{L}[A \omega_1(1-\omega_1)-c]}
{\fp \fx} \\
\frac{2 \sqrt{L}[A \omega_1(1-\omega_1)-c]}{\fp \fx} &
\frac{2[A(1-\omega_1)^2+c]}{\fx^2}
\label{massmatrix1}
\end{pmatrix} .
\ee
The eigenvalues of this matrix are:
\be
m^2_{S_1,S_2} = \frac{Z_L \mp \sqrt{Z_L^2-4Q_L}}{2} ,
\label{autovalori1}
\ee
where:
\bea
Z_L &\equiv& \frac{2A[\fp^2 (1-\omega_1)^2 + L \fx^2 \omega_1^2] +
2c(\fp^2+L\fx^2)}{\fp^2 \fx^2} , \nonumber \\
Q_L &\equiv& \frac{4LAc}{\fp^2 \fx^2} .
\label{zl-ql}
\eea
Making use of the following $N$--dependences of the relevant quantities
in the limit of large number of colours $N$ (see Ref. \cite{EM1994a}):
\be
\fp= \OO(N^{1/2}); \quad \fx= \OO(N^{1/2}); \quad A=\OO(1); \quad c=\OO(N) ,
\label{relazionin}
\ee
we derive, at the first order in the $1/N$ expansion (and {\it assuming} that
$c_1 \neq 0$: see the discussion in Appendix B), the following
expressions for the two eigenvectors:
\bea
S_1 &=& \frac{1}{\sqrt{\fp^2 + L\fx^2}} (\fp S_{\pi} + \sqrt{L} \fx S_X),
\nonumber \\
S_2 &=& \frac{1}{\sqrt{\fp^2 + L\fx^2}} (\sqrt{L} \fx S_{\pi} -\fp S_X),
\label{S1-S2}
\eea
with the corresponding eigenvalues:
\bea
m_{S_1}^2 &=& \frac{2LA}{\fp^2 + L \fx^2} = \OO(1/N) ,\nonumber \\
m_{S_2}^2 &=& \frac{2c(\fp^2 + L \fx^2)}{\fp^2 \fx^2} = \OO(1) .
\label{massaS1-S2}
\eea
The two fields $S_1$ and $S_2$ have the same quantum numbers, but different
quark contents: the first one (assuming that $\fp \gg \fx$) is prevalently
a quark--antiquark singlet $S_{\pi}$, while the second one is prevalently
an exotic $2L$--fermion singlet $S_X \sim i[\det(\bar{q}_{sL} q_{tR}) -
\det(\bar{q}_{sR}q_{tL})]$. Both fields are massive in the chiral limit.
If we let $F_X \to 0$ in the above--reported formulae (i.e., if we neglect the
new $U(1)$ axial condensate), then $S_1 \to S_{\pi}$ and $m^2_{S_1} \to
2LA / \fp^2$, which is the usual Witten--Veneziano formula for the $\eta'$
mass in the chiral limit \cite{Witten1979a,Veneziano1979}.
On the other side, $m_{S_2}^2 \simeq 2c/\fx^2 \to \infty$ for $\fx \to 0$,
being $c = \OO(\fx)$ [Eq. \eqref{c-mu}], and therefore, in this limit,
the field $S_2 \to -S_X$ is ``constrained'' to be zero.\footnote{More
rigorously, before
taking the limit $\fx \to 0$ (i.e., $X \to 0$), one should first take the limit
$\omega_1 \to 1$, so that no singular behaviour arises from the anomalous term
in Eqs. \eqref{lagrangian} and \eqref{newlagr} and the Lagrangian simply
reduces, for $X \to 0$, to the usual Lagrangian of Witten, Di Vecchia,
Veneziano {\it et al.} It is easy to check that, by putting $\omega_1 = 1$ in
Eqs. \eqref{autovalori1}--\eqref{zl-ql} and then letting $\fx \to 0$, one
recovers the same results that one also obtains by simply letting $\fx \to 0$
in Eqs. \eqref{massaS1-S2}, i.e., $m^2_{S_1} \to 2LA / \fp^2$ and
$m_{S_2}^2 \simeq 2c/\fx^2 \to \infty$.}
In the more general case $F_X \neq 0$, which we are considering in this paper,
there is a field ($S_1$) with a squared mass which vanishes as $\OO(1/N)$
in the large--$N$ expansion; on the contrary, the field $S_2$ has a {\it large}
mass of order $\OO(1)$ in the large--$N$ limit. It is quite easy to convince
oneself that the particle associated with the field $S_1$ is nothing but
the particle $\eta'$, which is required by the well--known
{\it Witten--Veneziano mechanism} for the solution of the {\it $U(1)$ problem}
(see Refs. \cite{EM1994c,EM2002}). In fact, the expression for the $U(1)$ axial
current:
\bea
J_{5, \mu}^{(L)} &=& i [\Tr(U^{\dagger} \partial_{\mu}U- U
\partial_{\mu}U^{\dagger}) + L (X^{\dagger} \partial_{\mu}X -
X \partial_{\mu}X^{\dagger})] \nonumber \\
&=& -\sqrt{2L} \partial_{\mu}(\fp S_{\pi} + \sqrt{L}F_X S_X) ,
\label{jmu5}
\eea
can be re--written, using the first Eq. \eqref{S1-S2}, as:
\be
J_{5, \mu}^{(L)}= - \sqrt{2L} F_{S_1} \demu S_1
\label{jmu5bis}
\ee
where :
\be
F_{S_1}= \sqrt{\fp^2 + L \fx^2}
\label{F_S1}
\ee
is nothing but the {\it decay constant} of the singlet meson $S_1$, defined as:
\be
\langle 0 | J_{5,\mu}^{(L)}(0)|S_1 (\vec{p}_1) \rangle =
i \sqrt{2L} F_{S_1} p_{1\mu} .
\ee
We remind that, according to the Witten--Veneziano mechanism for the solution
of the $U(1)$ problem, the $\eta'$ mass must satisfy the following relation,
known as the {\it Witten--Veneziano formula}:
\be
m_{\eta'}^2= \frac{2LA}{\fn^2}.
\label{relazionegenerale}
\ee
Using the first Eq. \eqref{massaS1-S2}, together with Eq. \eqref{F_S1},
one immediately verifies that the singlet meson associated with the field $S_1$
indeed verifies this relation, i.e., $m^2_{S_1}= 2LA/ F_{S_1}^2$.
For this reason, from now on, the field/particle $S_1$ will be denoted as
$\eta'$, with:
\be
\fn \equiv F_{S_1}= \sqrt{\fp^2 + L \fx^2} .
\ee
Instead, from now on, we shall use the name $\eta_X$ to denote the other
{\it exotic} singlet field/particle $S_2$.

\newsubsection{Mass spectrum at $T=0$ for the realistic $L=3$ case}
\label{spettrodimassaper}

\noindent
Let us consider more carefully the {\it realistic} case \cite{EM1994c},
in which there are $L=3$ light quark flavours, named $u$, $d$ and $s$, with
masses $m_u=(1.7\div3.3) \ \mev$, $m_d=(4.1\div5.8) \ \mev$ and $m_s=(80\div130)
\ \mev$ \cite{PDG2010}, which are small compared to the QCD mass--scale
$\Lambda_{QCD} \sim 0.5 \ \gev$.
In this case Eq. \eqref{Uexp} becomes:
\be
U= \sqrt{\frac{1}{2}} \fp \exp \left \{ \frac{i \sqrt{2}}{\fp}\Phi \right \},
~~~ \Phi= \sum_{a=1}^{8}{ \pi_a \tau_a + \frac{S_{\pi}}{\sqrt{3}} {\bf{I}} } ,
\label{Uexp3}
\ee
where $\pi_a \ (a=1, \cdots,8)$ are the pseudoscalar mesons ($J^P=0^-$) of the
octet, while $S_\pi$ is the quark--antiquark $SU(3)$--singlet field.
Proceeding as in the previous section, but making also an expansion up to the
first order in the quark masses, we immediately find that the fields
$\pi_1$, $\pi_2$, $\pi_4$, $\pi_5$, $\pi_6$, $\pi_7$ are already diagonal,
with masses:
\bea
m_{\pi_ {1,2}}^2 &\equiv& m^2_{\pi^ {\pm}} = B(m_u+m_d) , \nonumber \\
m_{\pi_{4,5}}^2 &\equiv& m^2_{K^{\pm}} = B(m_u+m_s) , \nonumber \\
m_{\pi_{6,7}}^2 &\equiv& m^2_{K^0,\bar{K}^0} = B(m_d+m_s) ,
\label{massemesoni}
\eea
where $B \equiv \frac{B_m}{2 \fp}$. \\
On the contrary the fields $\pi_3$, $\pi_8$, $S_{\pi}$, $S_X$ mix together,
with the following squared mass matrix:
\be
\mathcal{K}=
\begin{pmatrix}
2B\tilde{m} & \frac{1}{\sqrt{3}} B \Delta & \sqrt{\frac{2}{3}} B \Delta & 0 \\
\frac{1}{\sqrt{3}} B \Delta & \frac{2}{3} B (\tilde{m}+2m_s) &
\frac{2\sq}{3} B(\tilde{m}-m_s) & 0 \\
\sqrt{\frac{2}{3}} B \Delta & \frac{2\sq}{3} B(\tilde{m}-m_s) &
\frac{6 (A \omega_1^2+c)}{F_{\pi}^2} + m_0^2 &
\frac{2 \sqrt{3}[A(1-\omega_1)\omega_1-c]}{F_{\pi}F_X} \\
0 & 0 & \frac{2 \sqrt{3}[A(1-\omega_1)\omega_1-c]}{F_{\pi}F_X} &
\frac{2[A(1-\omega_1)^2+c]}{F_X^2}
\end{pmatrix} ,
\label{massmatrix2}
\ee
where $\tilde{m}\equiv \frac{m_u+m_d}{2}$, $m_0^2 \equiv \frac{2}{3}
B(2 \tilde{m}+m_s)$ and $\Delta \equiv m_u-m_d$. This last parameter $\Delta$
measures isospin violations, i.e., the explicit breaking of the $SU(2)_V$
symmetry. If we neglect the experimentally small violations of the $SU(2)_V$
isospin symmetry, i.e., if we put $\Delta=0$ in Eq.
\eqref{massmatrix2}\footnote{In the next section, instead, we shall take into
account also the small violations of the $SU(2)_V$ isospin symmetry, by taking
$\Delta \ne 0$.}, the squared mass matrix \eqref{massmatrix2} simplifies to:
\be
\mathcal{K}_0=
\begin{pmatrix}
2B\tilde{m} & 0 & 0 & 0 \\
0 & \frac{2}{3} B (\tilde{m}+2m_s) & \frac{2\sq}{3} B(\tilde{m}-m_s) & 0 \\
0 & \frac{2\sq}{3} B(\tilde{m}-m_s) & \frac{6 (A \omega_1^2+c)}{F_{\pi}^2} +
m_0^2 & \frac{2 \sqrt{3}[A(1-\omega_1)\omega_1-c]}{F_{\pi}F_X} \\
0 & 0 & \frac{2 \sqrt{3}[A(1-\omega_1)\omega_1-c]}{F_{\pi}F_X} &
\frac{2[A(1-\omega_1)^2+c]}{F_X^2}
\end{pmatrix} .
\label{massmatrix3}
\ee
Therefore, in this limit, $\pi_3$ also becomes diagonal and can be identified
with the physical state $\pi^0$, with squared mass:
\be
m_{\pi^0}^2 = 2 B \tilde{m} = B(m_u + m_d) \equiv m_{\pi}^2 .
\label{mass0}
\ee
The fields $(\pi_3,\pi_8,S_\pi,S_X)$ can be written in terms of the eigenstates
$(\pi^0,\eta,\eta',\eta_{X})$ as follows:
\be
\begin{pmatrix}
\pi_3 \\ \pi_8 \\ S_{\pi} \\ S_X
\end{pmatrix} = \mathbf{C}_0
\begin{pmatrix}
\pi^0 \\ \eta \\ \eta' \\ \eta_X
\end{pmatrix} ,
\label{autostati}
\ee
where $\mathbf{C}_0$ is the following orthogonal matrix \cite{EM2003,EM2004}:
\be
\mathbf{C}_0 =
\begin{pmatrix}
1 & 0 & 0 & 0 \\
0 & \cos\tilde{\varphi} & -\sin\tilde{\varphi} & 0 \\
0 & \frac{F_{\pi}}{F_{\eta'}} \sin\tilde{\varphi} &
\frac{F_{\pi}}{F_{\eta'}} \cos\tilde{\varphi} &
\frac{\sqrt{3}F_{X}}{F_{\eta'}} \\
0 & \frac{\sqrt{3} F_{X}}{F_{\eta'}} \sin\tilde{\varphi} &
\frac{\sqrt{3} F_{X}}{F_{\eta'}} \cos\tilde{\varphi} &
-\frac{F_{\pi}}{F_{\eta'}}
\end{pmatrix} .
\label{C0matrix}
\ee
As we have already said above,
$F_{\eta'} \equiv \sqrt{\csti}$ can be identified with the $\eta'$ decay
constant in the chiral limit of zero quark masses \cite{EM2003,EM2004}.
Moreover, $\tilde{\varphi}$ is a mixing angle, which can be related to the
masses of the quarks $m_u$, $m_d$, $m_s$, and therefore to the masses of the
octet mesons, by the relation (\ref{phitilde}) \cite{EM2003,EM2004}.\\
The matrix $\mathbf{C}_0$ has been derived by diagonalizing the squared mass
matrix \eqref{massmatrix3} at the first order in the quark masses and in
$1/N$, so neglecting terms behaving as $1/N^2$, $m^2$ or $m/N$ (and
{\it assuming}, again, that $c_1 \neq 0$: see the discussion in Appendix B).
Following Refs. \cite{Veneziano1979,DV1980,DNPV1981},
we have considered the limit in which $m/\Lambda_{QCD} \ll 1/N \ll 1$:
this particular choice is
justified by the fact that the mixing angle, which is of order
$\OO(mN/\Lambda_{QCD})$, is experimentally small\footnote{In the literature,
also other possibilities have been studied. For example, Leutwyler in Ref.
\cite{Leutwyler1996} considers $m/\Lambda_{QCD}$ and $1/N$ to be of the
same order, and Witten in Ref. \cite{Witten1980} studies also the opposite
case, i.e., $m N/\Lambda_{QCD} \gg 1$.}.
The other eigenvalues of the squared mass matrix \eqref{massmatrix3} can be
easily derived at the first order in the quark masses and in $1/N$ (in the
sense explained above):
\bea
\label{mass1}
m^2_{\eta} &=& \frac{2}{3}B(\tilde{m}+2m_s) , \\
\label{mass2}
m^2_{\eta'} &=& \frac{6A}{\fn^2} + \frac{\fp^2}{\fn^2} m_0^2 , \\
\label{mass3}
m^2_{\eta_X} &=& \frac{2c \fn^2}{\fp^2 \fx^2}
+ \frac{2A[\fp^2(\omega_1-1)+3\fx^2\omega_1]^2}{\fp^2 \fx^2 \fn^2}
+ \frac{3 \fx^2}{\fn^2} m_0^2 .
\eea
The physical interpretation of these three states is clear. The state $\eta$
is the eighth pseudo--Goldstone bosons of the octet: its mass vanishes with the
light quark masses. On the contrary, the states $\eta'$ and $\eta_X$ have
masses which do not vanish with the light quark masses. In particular, the
state $\eta'$ has a {\it topological} (non--chiral) squared mass term
$6A/F_{\eta'}^2$, which vanishes as $1/N$ in the large--$N$ limit.
The state $\eta_X$, instead, should be heavier, having a {\it normal} (non
chiral) {\it mesonic} mass term\footnote{See Ref. \cite{Witten1979b} for a
detailed discussion of hadrons and their masses in the framework of the $1/N$
expansion.} of order $\OO(1)$ in the large--$N$ limit.\\
From Eqs. \eqref{massemesoni}, \eqref{mass0} and \eqref{mass1} one
immediately derives the well--known {\it Gell-Mann--Okubo formula}
\cite{GellMann1961-1964,Okubo1961-1962} for the squared masses
of the octet mesons:
\be
3 m_{\eta}^2+m_{\pi}^2= 4 m_K^2,
\label{GMO}
\ee
where: $m_K^2 \equiv \frac{1}{2}(m^2_{K^{\pm}} + m^2_{K^0,\bar{K}^0})
= B(\tilde{m}+m_s)$.
In fact, it is natural to expect that the introduction of a new chiral order
parameter, which only breaks the $U(1)$ axial symmetry, should not modify
the mass relations for the octet mesons, such as Eq. \eqref{GMO}, which only
derive from the breaking of $SU(3)\otimes SU(3)$ down to $SU(3)_V$. \\
Considering also the squared mass \eqref{mass2} of the $\eta'$, one 
immediately derives the following interesting relation
(with $m_K^2$ defined as in Eq. \eqref{GMO}) \cite{EM1994c}:
\be
\left(1+ 3 \frac{\fx^2}{\fp^2} \right)m_{\eta'}^2 + m_{\eta}^2 - 2m_K^2
= \frac{6A}{\fp^2}.
\label{GMO2}
\ee
This is nothing but a generalization of the usual Witten--Veneziano formula
for the $\eta'$ mass (including non--zero quark masses), with a correction
which {\it only} depends on the parameter $\fx$ (which, as we have already
said in the Introduction, is essentially proportional to the new $U(1)$
axial condensate), but {\it not} on the other unknown parameters of the
model ($\omega_1$, $c_1$).
From Eq. \eqref{GMO2}, using the known values for the meson masses, the pion
decay constant $F_\pi$ and the pure--gauge topological susceptibility $A$,
one can derive the following upper limit for the parameter $F_X$:
$|F_X| \lesssim 20 \ \mev$ \cite{EM1994c,EM2002}. \\
Finally, we can derive an anologous relation involving also the squared mass
of the {\it exotic} state $\eta_X$. By taking the trace of the squared mass
matrix \eqref{massmatrix2}, using the relations \eqref{massemesoni}, together
with $\tilde{m}\equiv \frac{m_u+m_d}{2}$, $m_0^2 \equiv \frac{2}{3}
B(2 \tilde{m}+m_s)$ and $m_K^2 \equiv B(\tilde{m}+m_s)$, one obtains:
\bea
\Tr [\mathcal{K}] &=& m_{\pi^0}^2+m^2_{\eta}+m^2_{\eta'}+m^2_{\eta_X}
\nonumber \\
&=& 2B\tilde{m} + \frac{2}{3} B (\tilde{m}+2m_s) + m_0^2
+ \frac{6(A \omega_1^2+c)}{\fp^2}+ \frac{2A(1-\omega_1)^2+2c}{F_X^2}
\nonumber \\
&=& m^2_{\pi^0} + 2m^2_K + \frac{6(A \omega_1^2+c)}{\fp^2} +
\frac{2A(1-\omega_1)^2+2c}{F_X^2} ,
\label{GMO4}
\eea
from which, re--ordering, one finally gets:
\be
m^2_{\eta_X}+m^2_{\eta'}+m^2_{\eta}-2m^2_K =
\frac{2c \fn^2}{\fp^2 \fx^2}
+ \frac{2A[\fp^2(1-\omega_1)^2+3\fx^2\omega_1^2]}{\fp^2 \fx^2} .
\label{GMO3}
\ee
Unfortunately, this expression depends upon {\it all} the unknown parameters
of the model ($\fx$, $\omega_1$, $c_1$) and, therefore, we cannot use it to
obtain a direct estimate of the mass of the particle $\eta_X$.
However, in the next section we shall find some relations between its mass and
its decay widths, which in principle might be useful to identify a possible
candidate for this particle.

\newpage

\newsection{The strong decays of the pseudoscalar mesons $\eta,\eta',\eta_X$}

\noindent
In this section we shall study the strong decays of pseudoscalar mesons,
using the Chiral Effective Lagrangian which we have discussed above.\\
First we observe that the strong decays of a pseudoscalar meson into {\it two}
pseudoscalar mesons are trivially forbidden by parity conservation.
In fact, in terms of the Chiral Effective Lagrangian \eqref{newlagr}, one
easily verifies that it is invariant under the following field transformation:
\be
U \rightarrow U^{\dagger}, \quad X \rightarrow X^{\dagger},
\quad Q \rightarrow -Q,
\label{simmetrp}
\ee
which is nothing but the parity transformation for the fields [provided one
also transforms the space--time coordinates as $x = (x^0,\vec{x}) \to x_P =
(x^0,-\vec{x})$]. In terms of the meson fields $\pi_a,\ S_{\pi},\ S_X$,
defined in Eqs. \eqref{Uexp} and \eqref{Xexp}, it corresponds to:
\be
\pi_a \rightarrow -\pi_a \ ,\quad S_{\pi} \rightarrow -
S_{\pi}\ ,\quad S_{X}\rightarrow-S_{X}.
\label{parit}
\ee
Therefore, terms with an odd number of meson fields (which are not parity
invariant) necessarily vanish. In particular, operators with three pseudoscalar
meson fields are absent and therefore the strong decays of a pseudoscalar meson
into {\it two} pseudoscalar mesons are forbidden.\\
On the contrary, the strong decays of a pseudoscalar meson into {\it three}
pseudoscalar mesons, being induced by parity--invariant four--meson operators,
are allowed and we shall devote the rest of this section to a detailed
discussions of these decays.

\newsubsection{The four--meson Lagrangian}

\noindent
In order to study the strong decays of $\eta,~\eta',~\eta_X$ into three
pseudoscalar mesons, we have to isolate the four--meson operators in the
Lagrangian \eqref{newlagr}, when expanding the fields \eqref{Uexp} and
\eqref{Xexp} in powers of the meson fields. We thus obtain the following
{\it four--meson Lagrangian}:
\bea
\label{L4}
\La_4 &=& \frac{1}{4 F_{\pi}^2} \Tr \left[ \demu \Phi^2 \demup \Phi^2 +
\frac{4}{3} \Phi^3 \Box \Phi \right] + \frac{1}{4 F_{X}^2} \left[
\demu S_X^2 \demup S_X^2 + \frac{4}{3} S_X^3 \Box S_X \right] \nonumber \\
&+& \frac{B}{6 F_{\pi}^2} \Tr \left[ M \Phi^4\right] + \frac{c}{6}
\left( \frac{\sqrt{3}}{F_{\pi}} S_\pi - \frac{1}{F_X} S_X \right)^4 ,
\eea
where, as usual: $B= \frac{B_m}{2F_{\pi}}$,
$c= \frac{c_1}{\sqrt{2}} \left(\frac{F_X}{\sqrt{2}}\right)
\left(\frac{F_{\pi}}{\sqrt{2}}\right)^3$.\\
By making an integration by parts and using the usual identitites
for the $SU(3)$ generators, we can re--write the first term
in the r.h.s. of Eq. \eqref{L4} as (apart from total derivatives):
\bea
\delta\La_4^{(f)} &=& \frac{1}{4 F_{\pi}^2} \Tr \left[
\demu \Phi^2 \demup \Phi^2 + \frac{4}{3} \Phi^3 \Box \Phi \right] =
\frac{1}{4 F_{\pi}^2} \Tr \left[ \demu \Phi^2 \demup \Phi^2 -
\frac{4}{3} \demu \Phi^3 \demup \Phi \right] \nonumber \\
&=& \frac{1}{4 F_{\pi}^2} \left[ -\frac{2}{3}f_{ijc} f_{c \alpha \beta}
(\pi_i \demu \pi_j) (\pi_{\alpha}\demup \pi_{\beta}) \right],
\label{termine1}
\eea
where $f_{abc}$ are the structure constants of $SU(3)$, defined as: $[ \tau_a,
\tau_b ] = i\sq f_{abc} \tau_c$, with $\Tr (\tau_a \tau_b) = \delta_{ab}$.
It is easy to see that this term gives contributions only to decays into
charged pions, whose fields are $\pi^\pm = \frac{\pi_1 \mp i\pi_2}{\sqrt{2}}$.\\
Concerning the second term in the r.h.s. of Eq. \eqref{L4}, we immediately
recognize (after an integration by parts) that it vanishes (apart from a
total derivative):
\be
\frac{1}{4 F_{X}^2} \left[ \demu S_X^2 \demup S_X^2 +
\frac{4}{3} S_X^3 \Box S_X \right] =
\frac{1}{4 F_{X}^2} \left[ \demu S_X^2 \demup S_X^2 -
\frac{4}{3} \demu S_X^3 \demup S_X \right] = 0 .
\ee
Therefore, the four--meson Lagrangian \eqref{L4} reduces to:
\bea
\La_4 &=& \frac{1}{4 F_{\pi}^2} \left[ -\frac{2}{3}f_{ijc} f_{c \alpha \beta}
\pi_i \pi_\alpha \demu \pi_j \demup \pi_\beta \right] +
\frac{B}{6 F_{\pi}^2} \Tr \left[ M \Phi^4\right] \nonumber \\
&+& \frac{c}{6} \left( \frac{\sqrt{3}}{F_{\pi}} S_\pi -
\frac{1}{F_X} S_X \right)^4 .
\label{lag4}
\eea
In the limit $c \rightarrow 0$, $F_X \rightarrow 0$ and $S_X \rightarrow 0$
this Lagrangian reduces to the usual four--meson Lagrangian derived by
Di Vecchia {\it et al.} in Ref. \cite{DNPV1981}.\\
The last term in the four--meson Lagrangian \eqref{lag4} can be re--written
in terms of the mass eigenstates, given, in the case $\Delta = 0$, by
Eqs. \eqref{autostati}--\eqref{C0matrix}, so obtaining:
\be
\delta \La_4^{(c)} = \frac{c}{6} \left( \frac{\sqrt{3}}{F_{\pi}} S_\pi -
\frac{1}{F_X} S_X \right)^4
= \frac{c}{6}\left( \frac{\fn}{\fp \fx} \right)^4 \eta_X^4 .
\label{l4c}
\ee
This term contributes only to the elastic scattering amplitude $\eta_X \eta_X
\to \eta_X \eta_X$. At the end of the next subsection we shall see that, for
$\Delta \equiv m_u-m_d \neq 0$, the term $\delta\La_4^{(c)}$ gives also
contributions to the decays into three pseudoscalar mesons, but these
contributions are strongly suppressed for small $\Delta$.

\newsubsection{The mass eigenstates in the case $\Delta \neq 0$}

\noindent
In the strong decays of $\eta,\eta',\eta_X$ into three pions the $SU(2)$
isotopic spin $\hat{\vec{I}}$ is {\it not} conserved, i.e. (being the charge
conjugation $\hat{C}$ conserved by strong interactions) the so--called
$G$--{\it parity}, defined, for a multiplet of isotopic spin $I$, as
$\hat{G} \equiv \hat{C}e^{i\pi\hat{I}_2} = C_0 (-1)^I$, $C_0$ being the
eigenvalue of $\hat{C}$ for the neutral component of the multiplet,
is {\it not} conserved.
The mesons $\eta$, $\eta'$, $\eta_X$ are isosinglets ($I=0$) with $C=1$
(they can decay into $\gamma\gamma$ for the electromagnetic interaction),
and so they have $G=1$. On the contrary, the mesons $\pi$ form an isotriplet
($I=1$), with $C_0 = 1$ (since $\pi^0$ can decay into $\gamma\gamma$ for the
electromagnetic interaction), and so each of them has $G=-1$, and a
three--pion final state has $G = (-1)^3 = -1$.\\
We shall evaluate the decay amplitudes (and the corresponding decay widths)
at the lowest order in the parameter $\Delta \equiv m_u-m_d$, which measures
isospin violations, i.e., the explict breaking of the $SU(2)_V$ symmetry.
In the case $\Delta \neq 0$, the fields $\pi_{3},\pi_{8},S_{\pi},S_{X}$ mix
together with the squared mass matrix ${\mathcal{K}}$, given by Eq.
\eqref{massmatrix2}, while the remaining $\pi_a$ are already diagonal
\cite{EM1994c}.
We write the matrix ${\mathcal{K}}$ as:
\be
{\mathcal{K}}= {\mathcal{K}}_0 + \delta{\mathcal{K}}_{\Delta} ,
\ee
where ${\mathcal{K}}_0$ is the matrix ${\mathcal{K}}$ for $\Delta=0$, given
by Eq. \eqref{massmatrix3}, which is diagonalized by the orthogonal matrix
$\mathbf{C}_0$, given by Eq. \eqref{C0matrix}, while
$\delta{\mathcal{K}}_{\Delta}$ is given by:
\be
\delta{\mathcal{K}}_{\Delta}=
\begin{pmatrix}
0 & \frac{1}{\sqrt{3}} B \Delta & \sqrt{\frac{2}{3}} B \Delta & 0 \\
\frac{1}{\sqrt{3}} B \Delta & 0 & 0 & 0 \\
\sqrt{\frac{2}{3}} B \Delta & 0 & 0 & 0 \\
0 & 0 & 0 & 0
\end{pmatrix} .
\label{kdelta}
\ee
We shall evaluate the eigenvalues and the eigenstates of the matrix
${\mathcal{K}}$ at the first order in the parameter $\Delta$, by treating the
term $\delta{\mathcal{K}}_{\Delta}$ as a small perturbation. It is easy to
verify that the corrections to the eigenvalues (i.e., to the squared masses
$\mpi^2$, $\mn^2$, $\mnn^2$, $\mx^2$, evaluated in the previous section)
are of order $\Delta^2$ (the first--order corrections being identically zero)
and are therefore negligible, if we stop at the first order in $\Delta$.
Instead, the eigenstates of the matrix ${\mathcal{K}}$ at the first order
in the parameter $\Delta$ are given by:
\be
\begin{pmatrix}
\pi_3 \\ \pi_8 \\ S_{\pi} \\ S_X
\end{pmatrix} = \mathbf{C}
\begin{pmatrix}
\pi^0 \\ \eta \\ \eta' \\ \eta_X
\end{pmatrix},~~~~
\mathbf{C} =
\begin{pmatrix}
\delta_0 & \delta_1 & \delta_2 & \delta_3 \\
\alpha_0 & \alpha_1 & \alpha_2 & \alpha_3 \\
\beta_0 & \beta_1 & \beta_2 & \beta_3 \\
\gamma_0 & \gamma_1 & \gamma_2 & \gamma_3
\end{pmatrix},
\label{Cmatrix}
\ee
where:
\bea
\delta_0 &=& 1, \nonumber \\
\alpha_0 &=& \frac{B \Delta}{\sqrt{3}} \left[
\frac{\cos\tilde{\varphi}}{(\mpi^2 - \mn^2)}
\left( \cos\tilde{\varphi} + \frac{\sq \fp}{\fn} \sin\tilde{\varphi} \right)
\right. \nonumber \\
&-& \left. \frac{\sin\tilde{\varphi}}{(\mpi^2 - \mnn^2)}
\left( \frac{\sq \fp}{\fn} \cos\tilde{\varphi} - \sin\tilde{\varphi} \right)
\right] , \nonumber \\
\beta_0 &=& \frac{B \Delta}{\sqrt{3}} \left[
\frac{\fp \sin\tilde{\varphi}}{(\mpi^2 - \mn^2) \fn}
\left( \cos\tilde{\varphi} + \frac{\sq \fp}{\fn} \sin\tilde{\varphi} \right)
\right. \nonumber \\
&+& \left. \frac{\fp \cos\tilde{\varphi}}{(\mpi^2 - \mnn^2) \fn}
\left( \frac{\sq \fp}{\fn} \cos\tilde{\varphi} - \sin\tilde{\varphi} \right)
+ \frac{3\sq \fx^2}{(\mpi^2 - \mx^2) \fn^2} \right] , \nonumber \\
\gamma_0 &=& B \Delta \left[
\frac{\fx \sin\tilde{\varphi}}{(\mpi^2 - \mn^2) \fn}
\left( \cos\tilde{\varphi} + \frac{\sq \fp}{\fn} \sin\tilde{\varphi} \right)
\right. \nonumber \\
&+& \left. \frac{\fx \cos\tilde{\varphi}}{(\mpi^2 - \mnn^2) \fn}
\left( \frac{\sq \fp}{\fn} \cos\tilde{\varphi} - \sin\tilde{\varphi} \right)
- \frac{\sq \fp \fx}{(\mpi^2 - \mx^2) \fn^2} \right] , \nonumber \\
\delta_1 &=& \frac{B \Delta}{\sqrt{3} (\mn^2 - \mpi^2)}
\left( \cos\tilde{\varphi} + \frac{\sq \fp}{\fn} \sin\tilde{\varphi} \right) ,
\nonumber \\
\alpha_1 &=& \cos\tilde{\varphi} , \quad
\beta_1=\frac{\fp}{\fn} \sin\tilde{\varphi} , \quad
\gamma_1= \frac{\sqrt{3} F_X}{\fn} \sin\tilde{\varphi} ,
\nonumber \\
\delta_2 &=& \frac{B \Delta}{\sqrt{3} (\mnn^2 - \mpi^2)}
\left( \frac{\sq \fp}{\fn} \cos\tilde{\varphi} - \sin\tilde{\varphi} \right) ,
\nonumber \\
\alpha_2 &=& -\sin \tilde{\varphi} , \quad
\beta_2 = \frac{\fp}{\fn} \cos\tilde{\varphi} , \quad
\gamma_2 = \frac{\sqrt{3} F_X}{\fn} \cos\tilde{\varphi} , \nonumber \\
\delta_3 &=& \frac{\sq B \Delta \fx}{(\mx^2 - \mpi^2) \fn} ,\nonumber \\
\alpha_3 &=& 0, \quad \beta_3 = \frac{\sqrt{3}\fx}{\fn}, \quad
\gamma_3 = - \frac{\fp}{\fn} ,
\label{Cvalori}
\eea
where $\mpi^2$, $\mn^2$, $\mnn^2$, $\mx^2$ are given by Eqs. \eqref{mass0},
\eqref{mass1}, \eqref{mass2}, \eqref{mass3}.
The only modifications with respect to the ``unperturbed'' matrix
$\mathbf{C}_0$, reported in Eq. \eqref{C0matrix}, are in the elements
$\alpha_0$, $\beta_0$, $\gamma_0$, $\delta_1$, $\delta_2$, $\delta_3$,
which are now different from zero and of order $\Delta$:
in the limit $\Delta \rightarrow 0$ the matrix $\mathbf{C}$ correctly reduces
to the matrix $\mathbf{C}_0$.\\
At the end of the previous subsection we had observed that in the case
$\Delta=0$ the last term $\delta\La_4^{(c)}$ in the four--meson Lagrangian
\eqref{lag4}, being proportional to $\eta_X^4$,
contributes only to the elastic scattering amplitude $\eta_X \eta_X
\to \eta_X \eta_X$. Instead, in the realistic case in which
$\Delta \equiv m_u-m_d \neq 0$, this term has the form (obtained using Eqs.
\eqref{Cmatrix}--\eqref{Cvalori} derived above):
\be
\delta \La_4^{(c)} = \frac{c}{6} \left( \frac{\sqrt{3}}{F_{\pi}} S_\pi -
\frac{1}{F_X} S_X \right)^4
= \frac{c}{6} \left[ \left( \frac{\sq B \Delta}{(\mpi^2 - \mx^2) \fp} \right)
\pi^0 + \left( \frac{\fn}{\fp \fx} \right) \eta_X \right]^4 .
\label{cccc}
\ee
Therefore, when $\Delta\ne0$ this term contributes also to the decay
$\eta_X \rightarrow 3\pi^0$, but this contribution is of order $\OO(\Delta^3)$,
and therefore it is strongly suppressed, for small $\Delta$, when compared with
the similar contributions derived from the other terms in the Lagrangian
\eqref{lag4} [see Eq. \eqref{ampietax} below].
Therefore, in the following we shall neglect this contribution.

\newsubsection{Decays $\eta,\ \eta',\ \eta_X \rightarrow 3\pi^0, \
\pi^+ \pi^- \pi^0$}

\noindent
In this section we shall evaluate the {\it leading--order} (LO) amplitudes and
the corresponding widths for the decays of $\eta$, $\eta'$ and $\eta_X$ into
$3\pi^0$ or $\pi^+ \pi^- \pi^0$.
The fields in the four--meson Lagrangian $\La_4$, written in Eq. \eqref{lag4},
can be expressed in terms of the fields of the {\it physical} eigenstates
(which diagonalize the squared mass matrix $\mathcal{K}$) by using Eqs.
\eqref{Cmatrix}--\eqref{Cvalori}.\\
Let us start considering the decay $\eta \rightarrow 3 \pi^0$.
As we have already said after Eq. \eqref{termine1}, the first term (containing
field derivatives) of the four--meson Lagrangian $\La_4$ in Eq. \eqref{lag4}
does {\it not} contribute to this decay amplitude, which, therefore, turns
out to be simply a constant, i.e., {\it not} dependent on the particle momenta,
and given by, at the first order in the parameter $\Delta$:
\bea\lefteqn{
A(\eta \rightarrow 3\pi^0) = \left< \pi^0 \pi^0 \pi^0 | \La_4 | \eta \right> }
\nonumber \\
& & = \frac{B}{\sqrt{3} \fp^2} \left\{ \Delta (\alpha_1 + \sq \beta_1)
+ 2\sqrt{3}\tilde{m} \left[ \delta_1 +
(\alpha_1 + \sq \beta_1) (\alpha_0 + \sq \beta_0) \right] \right\} .
\label{ampi}
\eea
Using the expressions \eqref{Cvalori} for $\alpha_1,~\beta_1,~\delta_1,
~\alpha_0,~\beta_0$ and expanding up to the first order in the quark masses,
we obtain the following expression:
\be
A(\eta \rightarrow 3\pi^0) = \frac{B \Delta}{\sqrt{3} \fp^2} \left[
\cos \tilde{\varphi} + \frac{\sq \fp}{\fn} \sin \tilde{\varphi} \right] .
\label{ampiezza0}
\ee
The amplitude for the decay $\eta' \rightarrow 3 \pi^0$ can be obtained by
simply substituting $(\delta_1,~\alpha_1,~\beta_1)$ with
$(\delta_2,~\alpha_2,~\beta_2)$ in the expression \eqref{ampi}.
We thus obtain the following expression:
\be
A(\eta' \rightarrow 3\pi^0) = \frac{B \Delta}{\sqrt{3} \fp^2} \left[
\frac{\sq \fp}{\fn} \cos \tilde{\varphi} - \sin \tilde{\varphi} \right] ,
\label{ampipp}
\ee
Let us observe that in the limit $\fx \rightarrow 0$ (that is, $\fn \rightarrow
\fp$) the expressions \eqref{ampiezza0} and \eqref{ampipp} correctly reduce to
the corresponding expressions derived by Di Vecchia {\it et al.} in Ref.
\cite{DNPV1981}, i.e.:
\bea
A(\eta \rightarrow 3\pi^0)|_{F_X=0} &=& \frac{B \Delta}{\sqrt{3} \fp^2}
\left(\cos \varphi + \sqrt{2}\sin \varphi\right), \\
A(\eta' \rightarrow 3\pi^0)|_{F_X=0} &=& \frac{B \Delta}{\sqrt{3} \fp^2}
\left(\sq \cos\varphi - \sin\varphi \right),
\eea
where $\varphi$ is the mixing angle {\it without} the contribution coming from
the new $U(1)$ axial condensate, and it is given by Eq. \eqref{phitilde} with
$F_X=0$, i.e., $\tan\varphi = \frac{\sqrt{2}}{9A}B \fp^2 (m_s - \tilde{m}) =
\frac{\fp^2}{6 \sqrt{2}A} (m_{\eta}^2 - m_{\pi}^2)$.\\
From the amplitudes \eqref{ampiezza0} and \eqref{ampipp}, we can derive the
corresponding decay widths by integrating over the final--state phase space,
according to the formula (valid for {\it constant} amplitudes $A$ and three
{\it identical} final particles)
$\Gamma = \frac{1}{2M} \int \frac{1}{3!} \dd\Phi^{(3)} |A|^2 =
\frac{|A|^2}{2M \cdot 3!} \Phi^{(3)}$, where the total phase space
$\Phi^{(3)}$ is given by (see, for example, Ref. \cite{Davydychev2004}
and references therein):
\be
\Phi^{(3)} = \int\frac{\dd s \dd t}{128 \pi^3 M^2} = \frac{1}{128 \pi^3 M^2}
\int_{s_2}^{s_3}\frac{\dd s}{s} \sqrt{(s-s_1)(s-s_2)(s_3-s)(s_4-s)},
\label{integralone}
\ee
where $M$ is the mass of the initial particle and $s_1 \equiv (m_1-m_2)^2$,
$s_2 \equiv (m_1+m_2)^2$, $s_3 \equiv (M-m_3)^2$, $s_4 \equiv (M+m_3)^2$,
$m_1$, $m_2$ and $m_3$ being the masses of the three final particles;
$s$ and $t$ are the usual Mandelstam variables, defined as $s \equiv (P-p_1)^2$
and $t \equiv (P-p_2)^2$, where $P$ is the four--momentum of the initial
particle and $p_1$, $p_2$, $p_3$ are the four--momenta of the three final
particles ($P^2=M^2$, $p_1^2=m_1^2$, $p_2^2=m_2^2$, $p_3^2=m_3^2$). \\
After performing numerically the integration in Eq. \eqref{integralone} for the
two cases that we are considering, using the values for the meson masses as
reported by the \emph{Particle Data Group} \cite{PDG2010}, we have
obtained the following expressions for the decay widths:
\bea
\label{gamma0-eta}
\Gamma_{\rm LO} (\eta \rightarrow 3 \pi^0) &=& \frac{(B\Delta)^2}{36 \fp^4}
\left( \cos\tilde{\varphi} + \frac{\sq \fp}{\fn} \sin\tilde{\varphi} \right)^2
\frac{\Phi^{(3)}}{m_{\eta}} ,
~~~~\frac{\Phi^{(3)}}{m_{\eta}} = 9.82 \ \kev ,\\
\label{gamma0-eta'}
\Gamma_{\rm LO} (\eta' \rightarrow 3 \pi^0) &=& \frac{(B\Delta)^2}{36 \fp^4}
\left( \frac{\sq \fp}{\fn} \cos\tilde{\varphi} - \sin\tilde{\varphi} \right)^2
\frac{\Phi^{(3)}}{m_{\eta'}} ,
~~~~\frac{\Phi^{(3)}}{m_{\eta'}} = 67.00 \ \kev .
\eea
Proceeding analogously, the following expression is obtained for the 
leading--order amplitude of the decay $\eta_X \rightarrow 3 \pi^0$:
\be
A(\eta_X \rightarrow 3 \pi^0)=\frac{\sq B \Delta F_X}{\fp^2 \fn} .
\label{ampietax}
\ee
Let us observe that this amplitude correctly reduces to zero when
$\fx \rightarrow 0$, i.e., when the new $U(1)$ axial condensate is zero.
Concerning the derivation of the decay width, the mass $m_{\eta_X}$ of the
exotic meson $\eta_X$ is not directly known (but see the discussion at the end
of this subsection) and therefore the integration in Eq. \eqref{integralone}
cannot be performed numerically. However, on the basis of what we have said
in the previous section, the mass of the $\eta_X$ is expected to be quite
large, at least larger than the mass of the $\eta'$.
So it is probably not a too bad approximation to neglect the meson masses
in the total phase space for this process. In the limit $m_1 = m_2 = m_3 = 0$
Eq. \eqref{integralone} reduces to
\be
\Phi^{(3)}_0 (M) = \frac{M^2}{256 \pi^3},
\label{spaziof0}
\ee
and for the width of the decay $\eta_X \rightarrow 3\pi^0$ we obtain the
following approximate expression:
\be
\Gamma_{\rm LO} (\eta_X \rightarrow 3 \pi^0) = |A(\eta_X \rightarrow 3\pi^0)|^2
\frac{\Phi^{(3)}_0 (m_{\eta_X})}{2m_{\eta_X} \cdot 3!} =
\frac{(B\Delta)^2 F_X ^2}{1536 \pi^3 \fp^4 \fn^2} m_{\eta_X} .
\label{larghx1}
\ee
Let us now study the decays of $\eta$, $\eta'$ and $\eta_X$ into
$\pi^+ \pi^- \pi^0$. As we have already observed above, also the four--meson
Lagrangian term $\delta \La_4^{(f)}$, defined in Eq. \eqref{termine1} and
containing derivatives of the fields, gives contributions to the amplitudes
of these decays. In particular, one finds that:
\bea
\delta A^{(f)} (\eta \rightarrow \pi^+ \pi^- \pi^0) &=&
\left< \pi^+ \pi^- \pi^0 \left| \delta\La_4^{(f)} \right| \eta \right> =
\frac{1}{\fp^2} \delta_0 \delta_1 (s - s_0) , \\
\delta A^{(f)} (\eta' \rightarrow \pi^+ \pi^- \pi^0) &=&
\left< \pi^+ \pi^- \pi^0 \left| \delta\La_4^{(f)} \right| \eta' \right> =
\frac{1}{\fp^2} \delta_0 \delta_2 (s - s'_0) , \\
\delta A^{(f)} (\eta_X \rightarrow \pi^+ \pi^- \pi^0) &=&
\left< \pi^+ \pi^- \pi^0 \left| \delta\La_4^{(f)} \right| \eta_X \right> =
\frac{1}{\fp^2} \delta_0 \delta_3 (s - \bar{s}_0) ,
\eea
where the coefficients $\delta_0$, $\delta_1$, $\delta_2$ and $\delta_3$ are
defined in Eqs. \eqref{Cmatrix}--\eqref{Cvalori}, while $s_0$, $s'_0$ and
$\bar{s}_0$ are so defined:
\be
s_0 \equiv \frac{1}{3} (m_{\eta}^2 + 3 m_{\pi}^2) ,~~~
s'_0 \equiv \frac{1}{3} (m_{\eta'}^2 + 3 m_{\pi}^2) ,~~~
\bar{s}_0 \equiv \frac{1}{3} (m_{\eta_X}^2 + 3 m_{\pi}^2) ,
\ee
and, as usual, $s \equiv \left(P-P_{\pi^0} \right)^2 =
\left(P_{\pi^+}+P_{\pi^-} \right)^2$, $P$ being the four--momentum of the
initial particle ($\eta$, $\eta'$, $\eta_X$) and $P_{\pi^0}$, $P_{\pi^+}$,
$P_{\pi^-}$ being the four--momenta of the three final pions.\\
Adding also the contributions coming from the second term in the r.h.s. of
Eq. \eqref{lag4}, we obtain the following expressions for the amplitudes of
the decays $\eta,~\eta',~\eta_X \rightarrow \pi^+ \pi^- \pi^0$:
\bea
A(\eta \rightarrow \pi^+\pi^-\pi^0) &=& \frac{ B \Delta}{3 \sqrt{3} \fp^2}
\left(\cos \tilde{\varphi} + \frac{\sqrt{2}\fp}{\fn} \sin \tilde{\varphi}
\right) \left[ 1 + \frac{3(s-s_0)}{m_{\eta}^2-\mpi^2} \right] ,\\
A(\eta' \rightarrow \pi^+\pi^-\pi^0) &=& \frac{ B \Delta}{3 \sqrt{3} \fp^2}
\left( \frac{\sqrt{2}\fp}{\fn} \cos \tilde{\varphi} - \sin \tilde{\varphi}
\right) \left[ 1 + \frac{3(s-s'_0)}{m_{\eta'}^2-\mpi^2} \right] ,\\
A(\eta_X \rightarrow \pi^+ \pi^- \pi^0) &=&
\frac{\sq B \Delta F_X}{3 \fp^2 \fn}
\left[ 1 + \frac{3(s-\bar{s}_0)}{m_{\eta_X}^2-\mpi^2} \right] .
\eea
From these amplitudes we can derive the corresponding decay widths by
integrating over the final--state phase space, according to the formula
(see, for example, Ref. \cite{Davydychev2004} and references therein):
\bea
\Gamma &=& \frac{1}{2M} \int \dd \Phi^{(3)} |A|^2 =
\frac{1}{2M} \int \frac{\dd s \dd t}{128 \pi^3 M^2} |A|^2 \nonumber \\
&=& \frac{1}{256 \pi^3 M^3} \int_{s_2}^{s_3} \frac{\dd s}{s}|A(s)|^2
\sqrt{(s-s_1)(s-s_2)(s_3-s)(s_4-s)} ,
\label{integralone-bis}
\eea
where the notation is the same already used in Eq. \eqref{integralone}.
After performing numerically the integration in Eq. \eqref{integralone-bis},
using the values for the meson masses as reported by the \emph{Particle Data
Group} \cite{PDG2010}, we have obtained the following expressions for
the decay widths:
\bea
\label{gamma1-eta}
\Gamma_{\rm LO} (\eta \rightarrow \pi^+ \pi^- \pi^0) &=&
\frac{(B\Delta)^2}{54 \fp^4}
\left( \cos\tilde{\varphi} + \frac{\sq \fp}{\fn} \sin\tilde{\varphi} \right)^2
\times 10.48 \ \kev ,\\
\label{gamma1-eta'}
\Gamma_{\rm LO} (\eta' \rightarrow \pi^+ \pi^- \pi^0) &=&
\frac{(B\Delta)^2}{54 \fp^4}
\left( \frac{\sq \fp}{\fn} \cos\tilde{\varphi} - \sin\tilde{\varphi} \right)^2
\times 83.95 \ \kev .
\eea
Concerning the case of the decay $\eta_X \rightarrow \pi^+ \pi^- \pi^0$, we
proceed exactly as for the case of the decay $\eta_X \rightarrow 3\pi^0$ and
we neglect the meson masses in the calculation of the integral
\eqref{integralone-bis}, so obtaining the following approximate expression
for the decay width:
\be
\Gamma_{\rm LO} (\eta_X \rightarrow \pi^+ \pi^- \pi^0) =
\frac{(B\Delta)^2 F_X^2}{1536 \pi^3 \fp^4 \fn^2} m_{\eta_X} .
\ee
We now numerically compute our theoretical expressions for the leading--order
decay widths, using for the mixing angle $\tilde\varphi$ the value derived
from Eq. \eqref{phitilde}.\\
All our isospin--violating decay widths are proportional to the factor:
\be
(B \Delta)^2 = m_{\pi}^4 \left( \frac{m_u-m_d}{m_u+m_d} \right)^2
= m_{\pi}^4 \left( \frac{R-1}{R+1} \right)^2
\simeq 2.66 \times 10^7 \ \mev^4 ,
\ee
where $\mpi^2 = B(m_u+m_d) \simeq (134.98 \ \mev)^2$ and $R \equiv m_u/m_d
\simeq 0.558$ is the ratio between the {\it up} and {\it down} quark masses,
determined using Eqs. \eqref{massemesoni} and the experimental values of the
meson masses reported in the \emph{Particle Data Group} \cite{PDG2010}.\\
We are, of course, particularly interested in the effects due to a non--zero
value of the parameter $\fx$, related to the new $U(1)$ axial condensate
considered in this paper. In Table 1 we report, for each decay process
of the form $\eta,\eta' \rightarrow 3\pi$, the leading--order theoretical
prediction, using for the parameter $\fx$ the value $\fx=24(7)$ MeV, that
we have found studying the radiative decays $\eta,\eta'\rightarrow\gamma\gamma$
[see the Introduction and, in particular, Eq. \eqref{results}]. These values
are compared with the corresponding values obtained for $\fx=0$, i.e., in the
absence of the new $U(1)$ axial condensate (in Table 1 we also explicitly
show the {\it correction} to the decay widths, $\Delta\Gamma_{\rm LO} \equiv
\Gamma_{\rm LO}(\fx=24 \pm 7 ~{\rm MeV}) -\Gamma_{\rm LO}(\fx=0)$, coming from
a non--zero value of $\fx$), and also with the experimental values.

\begin{table}[t]
\centering
\begin{tabular}[h]{l|c|cc|c}
\hline
Decay & $\Gamma_{\rm exp}$ (keV) & \multicolumn{2}{c|}{$\Gamma_{\rm LO}$
(keV)} & $\Delta\Gamma_{\rm LO}$ (keV) \\
& & $F_X=0$ & $F_X=24(7)$ MeV & \\
\hline
\hline
& & & & \\
$\eta \rightarrow 3\pi^0$ & 0.423(26) & 0.178 & $0.176(1)$ & $-0.002(1)$ \\
$\eta' \rightarrow 3\pi^0$ & 0.33(6) & 0.84 & $0.62(10)$ & $-0.24(10)$ \\
$\eta \rightarrow \pi^+\pi^-\pi^0$ & 0.30(2) & 0.127 & $0.125(1)$ &
$-0.002(1)$ \\
$\eta' \rightarrow \pi^+\pi^-\pi^0$ & 0.70(25) & 0.70 &
$0.52(8)$ & $-0.18(8)$ \\
& & & & \\
\hline
\end{tabular}
\caption{The leading--order theoretical predictions for the decay widths,
computed both for $\fx=0$ and for $\fx=24(7)$ MeV, and the corresponding
{\it corrections} to the decay widths, $\Delta\Gamma_{\rm LO} \equiv
\Gamma_{\rm LO}(\fx=24 \pm 7 ~{\rm MeV}) -\Gamma_{\rm LO}(\fx=0)$,
compared with the experimental values.}
\label{tab1}
\end{table}

Concerning the comparison with the experimental values, it is well known that,
because of large unitarity corrections due to strong final--state interactions,
one has to go beyond leading and even one--loop order in chiral perturbation
theory in order to obtain a valid, reliable representation of the $\eta,\eta'
\rightarrow 3\pi$ decay amplitudes and of the corresponding decay widths, that
can be successfully compared with the experimental values
\cite{GL1985,KWW1996,AL1996,BB2003,BN2005,BG2007}.\\
In the present paper we are not, of course, aiming at that. In particular,
we cannot proceed as in the case of the radiative decays $\eta,\eta'\rightarrow
\gamma\gamma$, i.e., we cannot extract the value of $\fx$ (and of the mixing
angle $\tilde\varphi$) by comparing, e.g., the leading--order theoretical
predictions \eqref{gamma0-eta} and \eqref{gamma0-eta'}, for the
$\eta \rightarrow 3\pi^0$ and $\eta' \rightarrow 3\pi^0$ decay widths,
with the corresponding experimental values reported in Table 1.
Indeed, making use of Eq. \eqref{phitilde} for $\tan\tilde\varphi$, one easily
verifies that, being $\tan\tilde\varphi$ and $\fn \equiv \sqrt{\fp^2 + 3\fx^2}$
increasing functions of $\fx$, the expression \eqref{gamma0-eta} for
$\Gamma_{\rm LO}(\eta\rightarrow 3\pi^0)$ is a decreasing functions of $\fx$:
so, being its value at $\fx=0$ already smaller than the corresponding
experimental value, it turns out that there is no value of $\fx$ which
makes the expression \eqref{gamma0-eta} compatible with the experimental value
in Table 1.\footnote{Even considering the singlet decay constant $\fn$ and the
mixing angle $\tilde\varphi$ in Eqs. \eqref{gamma0-eta} and \eqref{gamma0-eta'}
as free parameters, to be fixed from a comparison with the experimental values
reported in Table 1, we would find a too small value $\fn \simeq 68$ MeV for
the singlet decay constant, incompatible with the formula \eqref{F-eta'}, i.e.,
$\fn = \sqrt{\fp^2 + 3\fx^2} \ge \fp = 92.2(4)$ MeV, and also an anomalously
large value $\tilde\varphi \simeq 44^\circ$ for the mixing angle.}\\
Instead, our aim is simply to quantify the corrections coming from a non--zero
value of the parameter $\fx$, taking the leading--order amplitudes/widths in the
$\fx=0$ case as a useful reference point. From the values reported in Table 1
we can conclude that:
\begin{itemize}
\item[i)] In the case of the $\eta \rightarrow 3\pi$ decays, the size of
the corrections $\Delta\Gamma_{\rm LO}$ coming from a non--zero value $\fx =
24(7)$ MeV, with respect to the $\fx=0$ case, is very small, being of the
order of $1\%$, i.e., comparable to (or even smaller than) the size of the
{\it electromagnetic} corrections for these decays, which have been recently
re--calculated in Ref. \cite{DKM2009}.
\item[ii)] Instead, in the case of the $\eta' \rightarrow 3\pi$ decays, the
size of the corrections $\Delta\Gamma_{\rm LO}$ is much larger, being of the
order of $30\%$. Moreover, at least for the decay $\eta' \rightarrow 3\pi^0$
(where the statistical errors are smaller), this (negative) correction seems
to go in the right direction, improving the agreement between the theoretical
prediction and the experimental value.
\end{itemize}

Concerning the decays of the $\eta_X$ into three pions, we derive the
following relations between its mass $m_{\eta_X}$ and the decay widths:
\be
\frac{\Gamma_{\rm LO} (\eta_X \rightarrow 3 \pi^0)}{m_{\eta_X}} =
\frac{\Gamma_{\rm LO} (\eta_X \rightarrow \pi^+ \pi^- \pi^0)}{m_{\eta_X}} =
(4.35^{+2.17}_{-1.97}) \times 10^{-7} .
\label{constraints}
\ee
These constraints could be used to identify a possible candidate for the
exotic singlet meson $\eta_X$, once we know its mass and decay widths.
According to the \emph{Particle Data Group} \cite{PDG2010}, the possible
candidates for the $\eta_X$, having the same quantum numbers
$I^G(J^{PC})=0^+(0^{-+})$ of the $\eta'$, but a larger mass, are the following:
\bea
\eta(1295)&:& \quad \Gamma_{\rm tot}= 55(5)\ \mev ,\nonumber \\
\eta(1405)&:& \quad \Gamma_{\rm tot}= 51(3)\ \mev ,\nonumber \\
\eta(1475)&:& \quad \Gamma_{\rm tot}= 85(9)\ \mev ,\nonumber \\
\eta(1760)&:& \quad \Gamma_{\rm tot}= 96(70)\ \mev ,\nonumber \\
\eta(2225)&:& \quad \Gamma_{\rm tot}= 185^{+70}_{-40} \ \mev .
\label{candidates}
\eea
Unfortunately, no quantitative determination of their decay widths into
three pions has been done up to now.

\newsubsection{Decays $\eta' \rightarrow \eta \pi \pi$ and
$\eta_X \rightarrow \eta \pi \pi, \ \eta' \pi \pi$}

\noindent
We now study the decays of $\eta'$ into $\eta \pi^0 \pi^0$,
$\eta \pi^+ \pi^-$ and of $\eta_X$ into $\eta \pi^0 \pi^0$, $\eta \pi^+ \pi^-$,
$\eta' \pi^0 \pi^0$, $\eta' \pi^+ \pi^-$.
This decays do not violate isospin and so they can happen also when $\Delta=0$.
Therefore, in order to evaluate the amplitudes and the corresponding widths
for these decays, we shall use the approximate expressions \eqref{C0matrix}
of the eigenstates at the order {\it zero} in the isospin--violating
parameter $\Delta$.\\
The following expression is obtained for the leading--order amplitudes of the
decays $\eta' \rightarrow \eta \pi^0 \pi^0$ and
$\eta' \rightarrow \eta \pi^+ \pi^-$
(which, in the limit of {\it exact} $SU(2)_V$ isospin symmetry, are equal):
\bea\lefteqn{
A(\eta' \rightarrow \eta \pi^0 \pi^0) =
A(\eta' \rightarrow \eta \pi^+ \pi^-) } \nonumber \\
& & = \frac{m^2_{\pi}}{6 \fp^2}\left[
\frac{2 \sq \fp}{\fn} \cos(2\tilde{\varphi}) +
\left(\frac{2\fp^2}{\fn^2}-1\right) \sin(2\tilde{\varphi}) \right] .
\label{ampiezzacost}
\eea
In the limit $\fx \to 0$ these amplitudes reduce to the expression already
found in Ref. \cite{DNPV1981}, i.e.:
\be
A(\eta' \rightarrow \eta \pi^0 \pi^0)|_{\fx = 0} =
A(\eta' \rightarrow \eta \pi^+ \pi^-)|_{\fx = 0} = \frac{m^2_{\pi}}{6 \fp^2}
\left[ 2\sqrt{2} \cos(2\varphi) + \sin(2\varphi) \right] .
\ee
After numerical integration of the phace space \eqref{integralone}, using the
values for the meson masses reported in Ref. \cite{PDG2010}, we obtain
the corresponding decay widths:
\bea
\Gamma_{\rm LO} (\eta' \rightarrow \eta \pi^0 \pi^0) &=&
\left|A(\eta' \rightarrow \eta \pi^0 \pi^0)\right|^2
\frac{\Phi^{(3)}}{2 m_{\eta'} \cdot 2!} ,
~~~~\frac{\Phi^{(3)}}{2 m_{\eta'} \cdot 2!} = 1.093 \ \kev ,\nonumber\\
\Gamma_{\rm LO} (\eta' \rightarrow \eta \pi^+ \pi^-) &=&
\left|A(\eta' \rightarrow \eta \pi^+ \pi^-)\right|^2
\frac{\Phi^{(3)}}{2 m_{\eta'}} =
2 \Gamma_{\rm LO} (\eta' \rightarrow \eta \pi^0 \pi^0) .
\eea
We proceed as in the previous subsection and numerically compute our
theoretical expressions for the leading--order decay widths, using for the
mixing angle $\tilde\varphi$ the value derived from Eq. \eqref{phitilde}
and for the parameter $\fx$ the value $\fx=24(7)$ MeV, that
we have found studying the radiative decays $\eta,\eta'\rightarrow\gamma\gamma$.
Again, our aim is simply to quantify the corrections coming from a non--zero
value of the parameter $\fx$, taking the leading--order amplitudes/widths in the
$\fx=0$ case as a reference point. In this case, however, it is already known
from Ref. \cite{DNPV1981} that the leading--order theoretical predictions
for $\fx=0$,
\be
\Gamma_{\rm LO} (\eta' \rightarrow \eta \pi^+ \pi^-)|_{\fx=0} =
2 \Gamma_{\rm LO} (\eta' \rightarrow \eta \pi^0 \pi^0)|_{\fx=0} = 2.42 \ \kev ,
\ee
are in {\it strong} disagreement with the experimental values \cite{PDG2010},
$\Gamma_{\rm exp} (\eta' \rightarrow \eta \pi^+ \pi^-) = 84(5)$ keV and
$\Gamma_{\rm exp} (\eta' \rightarrow \eta \pi^0 \pi^0) = 42(4)$ keV.\\
We can try to see if the introduction of a non--zero value of $\fx$ can cure,
at least in part, the strong disagreement between leading--order theoretical
predictions and experimental values: however, the answer to this question
is negative. In fact, we find that:
\be
\Gamma_{\rm LO} (\eta' \rightarrow \eta \pi^+ \pi^-)|_{\fx=24(7) \ \mev} =
2 \Gamma_{\rm LO} (\eta' \rightarrow \eta \pi^0 \pi^0)|_{\fx=24(7) \ \mev} =
1.78(30) \ \kev .
\ee
Even if the correction $\Delta\Gamma_{\rm LO}$ is quite large (of the order
of $30\%$) if compared with the value of $\Gamma_{\rm LO}$ at $\fx=0$,
it is, however, too small if compared with the experimental value.
In addition, the correction $\Delta\Gamma_{\rm LO}$, being negative, goes
in the ``wrong'' direction, lowering the theoretical prediction at $\fx=0$,
which is already much smaller than the experimental value:
in other words, it is not possible to find a value of the parameter $\fx$
which moves the leading--order theoretical prediction towards the experimental
value. Moreover, the amplitude \eqref{ampiezzacost} is a constant, while
the experimental data are well fitted by a non--constant amplitude having
the form: $A(\eta' \rightarrow \eta \pi \pi)= A(1- \sigma_1 T_{\eta})$,
where $T_{\eta}$ is the kinetic energy of the $\eta$, $A$ and $\sigma_1$
are some constants. As already observed in Ref. \cite{DNPV1981},
in order to describe this behaviour, and to obtain a better agreement with
the experimental value of the decay width, it is not enough to retain only
the leading order in the $1/N$ expansion, but one has to go to the
next--to--leading order, adding to the Lagrangian \eqref{lagrangian}
non--leading terms such as $\lambda Q^2 \Tr (\partial_\mu U \partial^\mu
U^\dagger)$, that may be very important because of the proportionality of
the leading terms to the tiny pion mass.\footnote{A different and alternative
approach, first suggested in Ref. \cite{DT1978}, considers the decay
$\eta' \rightarrow \eta\pi\pi$ to be dominated by coupling to nearby scalar
resonances.}
The systematic introduction, in our model, of higher--order terms in the $1/N$
expansion (including also one--loop graphs, which are of order $1/N^2$: see,
for example, Refs. \cite{GL1985,Leutwyler1998}) is, of course, a quite hard
task, which is beyond the aim of the present paper (but it will probably be
addressed in a subsequent work).

Concerning the exotic meson $\eta_X$, the following expressions are obtained
for the leading--order amplitudes of the decays
$\eta_X \rightarrow \eta \pi \pi$ and $\eta_X \rightarrow \eta' \pi \pi$:
\bea
A(\eta_X \rightarrow \eta \pi^0 \pi^0) &=&
A(\eta_X \rightarrow \eta \pi^+ \pi^-) =
\frac{\sq m_{\pi}^2 F_X}{\sqrt{3} \fp^2 \fn} \left( \cos\tilde{\varphi}
+ \frac{\sq \fp}{\fn} \sin\tilde{\varphi} \right), \\
A(\eta_X \rightarrow \eta' \pi^0 \pi^0) &=&
A(\eta_X \rightarrow \eta' \pi^+ \pi^-) =
\frac{\sq m_{\pi}^2 F_X}{\sqrt{3} \fp^2 \fn} \left( \frac{\sq \fp}{\fn}
\cos\tilde{\varphi} - \sin\tilde{\varphi} \right) .
\eea
From these amplitudes we can obtain the corresponding decay widths, using,
for the integrated phase space \eqref{integralone}, the following approximate
expression obtained neglecting the pion masses (while retaining the $\eta$ and
$\eta'$ masses different from zero):
\be
\Phi^{(3)}_1 (M,m) = \frac{M^4 - m^4 + 4 M^2 m^2 \ln (m/M)}{256 \pi^3 M^2} ,
\label{spaziofm}
\ee
where $M$ is the mass of the initial particle and $m$ is the mass of the
final massive particle. We thus find the following expressions:
\bea
\lefteqn{
\Gamma_{\rm LO} (\eta_X \rightarrow \eta \pi^0 \pi^0) =
\left|A(\eta_X \rightarrow \eta \pi^0 \pi^0)\right|^2
\frac{\Phi^{(3)}_1 (m_{\eta_X},m_\eta)}{2 m_{\eta_X} \cdot 2!} }
\nonumber \\
& & = \frac{\mpi^4 F_X^2}{1536 \pi^3 \fp^4 \fn^2}
\left( \cos\tilde{\varphi} + \frac{\sq \fp}{\fn} \sin\tilde{\varphi} \right)^2
\left[ \mx - \frac{\mn^4}{\mx^3} + \frac{4 \mn^2}{\mx} \ln \left(
\frac{\mn}{\mx} \right) \right]
\nonumber \\
& & = (0.95^{+0.46}_{-0.42}) \times 10^{-5} \mx
\left[ 1 - \left(\frac{\mn}{\mx}\right)^4 + 4\left(\frac{\mn}{\mx}\right)^2
\ln \left(\frac{\mn}{\mx}\right) \right] ,
\nonumber \\
\lefteqn{
\Gamma_{\rm LO} (\eta_X \rightarrow \eta \pi^+ \pi^-) =
\left|A(\eta_X \rightarrow \eta \pi^+ \pi^-)\right|^2
\frac{\Phi^{(3)}_1 (m_{\eta_X},m_\eta)}{2 m_{\eta_X}} }
\nonumber \\
& & = 2 \Gamma_{\rm LO} (\eta_X \rightarrow \eta \pi^0 \pi^0) ,
\eea
and also:
\bea
\lefteqn{
\Gamma_{\rm LO} (\eta_X \rightarrow \eta' \pi^0 \pi^0) =
\left|A(\eta_X \rightarrow \eta' \pi^0 \pi^0)\right|^2
\frac{\Phi^{(3)}_1 (m_{\eta_X},m_{\eta'})}{2 m_{\eta_X} \cdot 2!} }
\nonumber \\
& & = \frac{\mpi^4 F_X^2}{1536 \pi^3 \fp^4 \fn^2}
\left( \frac{\sq \fp}{\fn} \cos\tilde{\varphi} - \sin\tilde{\varphi} \right)^2
\left[ \mx - \frac{\mnn^4}{\mx^3} + \frac{4 \mnn^2}{\mx} \ln \left(
\frac{\mnn}{\mx} \right) \right]
\nonumber \\
& & = (0.49^{+0.12}_{-0.18}) \times 10^{-5} \mx
\left[ 1 - \left(\frac{\mnn}{\mx}\right)^4 + 4\left(\frac{\mnn}{\mx}\right)^2
\ln \left(\frac{\mnn}{\mx}\right) \right] ,
\nonumber \\
\lefteqn{
\Gamma_{\rm LO} (\eta_X \rightarrow \eta' \pi^+ \pi^-) =
\left|A(\eta_X \rightarrow \eta' \pi^+ \pi^-)\right|^2
\frac{\Phi^{(3)}_1 (m_{\eta_X},m_{\eta'})}{2 m_{\eta_X}} }
\nonumber \\
& & = 2 \Gamma_{\rm LO} (\eta_X \rightarrow \eta' \pi^0 \pi^0) .
\eea
As in the case of Eq. \eqref{constraints}, also these relations could in
principle be used to identify a possible candidate for the
exotic singlet meson $\eta_X$, once we know its mass and decay widths.
However, a certain caution must be used since, as in the case of the decays
$\eta' \rightarrow \eta \pi \pi$, large corrections to these leading--order
results could come from non--leading terms in the $1/N$ expansion: only a
detailed analysis of our model at the next--to--leading order in $1/N$
shall clarify this point.

\newsubsection{Possible decays $\eta_X \rightarrow 3 \eta, \ \eta \eta \eta',
\ \eta \eta'\eta', \ 3\eta'$ ?}

\noindent
If the exotic singlet meson $\eta_X$ were heavy enough, let us say, if
$\mx > 3\mn \simeq 1640$ MeV, it could also decay into three $\eta$ particles.
The amplitude for this decay, which does not violate $SU(2)_V$ isospin,
can be evaluated at the order {\it zero} in the
isospin--violating parameter $\Delta$, so using the approximate form
\eqref{C0matrix} for the physical eigenstates, and the following result
is obtained:
\be
A(\eta_X \rightarrow 3\eta) =
\frac{8\sq m^2_K \fx}{3 \sqrt{3} \fp^2 \fn} \left(
-\cos\tilde{\varphi} + \frac{3 \sq \fp}{2\fn} \sin\tilde{\varphi} \right) .
\ee
In order to estimate the decay width, considering that also the final--state
particles $\eta$ are rather heavy, we use the approximate expression for the
total phase space \eqref{integralone} in the {\it non--relativistic} limit,
i.e.:
\be
\Phi^{(3)}_{nr}(M,m_1,m_2,m_3) =
\frac{Q^2}{64 \pi^2}\sqrt{\frac{m_1 m_2 m_3}{(m_1+m_2+m_3)^3}} ,
\label{nonrel}
\ee
where $Q \equiv M-m_1-m_2-m_3$ is the so--called ``\emph{Q value}''
of the decay.\\
We thus obtain the following approximate expression for the decay width:
\bea
\lefteqn{
\Gamma_{\rm LO} (\eta_X \rightarrow 3 \eta) =
\left|A(\eta_X \rightarrow 3 \eta) \right|^2
\frac{\Phi^{(3)}_{nr}(\mx,\mn,\mn,\mn)}{2\mx \cdot 3!} }
\nonumber \\
& & = \frac{m^4_K \fx^2}{486 \sqrt{3} \pi^2 \fp^4 \fn^2} \left(
\cos\tilde{\varphi} - \frac{3 \sq \fp}{2\fn} \sin\tilde{\varphi} \right)^2
\frac{(\mx - 3\mn)^2}{\mx}
\nonumber \\
& & = (0.96^{+0.46}_{-0.43}) \times 10^{-3} \mx
\left( 1 - \frac{3\mn}{\mx} \right)^2 ,
\eea
where, as usual, we have used for the parameter $\fx$ the value $\fx = 24(7)$
MeV, that we have found studying the radiative decays
$\eta,\eta' \rightarrow \gamma\gamma$, and for the mixing angle $\tilde\varphi$
the value derived from Eq. \eqref{phitilde}. (For example, for a value
$\mx \approx 2$ GeV, one would get
$\Gamma_{\rm LO} (\eta_X \rightarrow 3 \eta) \approx 61$ keV.)\\
Other possible decays of this kind (supposing that the $\eta_X$ is heavy
enough so that they are kinematically allowed) are
$\eta_X \rightarrow \eta \eta \eta',\ \eta \eta' \eta', \ 3\eta'$,
and their amplitudes and corresponding widths can be derived in a similar way.

\newsection{Conclusions}

\noindent
In this paper we have considered a scenario (supported by some lattice results)
in which a $U(1)$--breaking condensate survives across the chiral transition
at $T_{ch}$, staying different from zero up to $T_{U(1)} > T_{ch}$.
This scenario has important consequences for the pseudoscalar--meson sector,
which can be studied using an effective Lagrangian model, including also
the new $U(1)$ chiral condensate. This model, originally proposed in
Refs. \cite{EM1994a,EM1994b,EM1994c,EM1995} and elaborated in Refs.
\cite{EM2002,EM2003,EM2004}, could perhaps be verified in the near future by
heavy--ion experiments, by analysing the pseudoscalar--meson spectrum
in the singlet sector.

Section 2 contains a brief review (for the benefit of the reader) of the main
results, obtained in the original papers \cite{EM1994a,EM1994c,EM2002},
concerning the mass spectrum of the Chiral Effective Lagrangian.
The Lagrangian \eqref{newlagr} contains a new field $X$ and three new
parameters, namely $F_X$, $\omega_1$ and $c_1$, with respect to the usual
Lagrangian of Witten, Di Vecchia, Veneziano {\it et al.} In this paper we have
{\it assumed} that the parameter $F_X$, which is essentially proportional to
the new $U(1)$ axial condensate, is different from zero.
In this case, there are two singlet pseudoscalar mesons, the $\eta'$ and an
{\it exotic} particle $\eta_X$, whose squared masses (assuming also that the
coupling constant $c_1$ of the interaction term
$\det(U)X^{\dagger} + \det(U^{\dagger})X$ in Eq. \eqref{potential},
between the usual $q\bar{q}$ meson field $U$ and the exotic meson field $X$,
is different from zero and not too small: see the discussion in Appendix B)
are given by Eqs. \eqref{mass2} and \eqref{mass3}:
in particular, the {\it exotic} particle $\eta_X$ turns out to have a
{\it large} (non--chiral) mass term of order $\OO(1)$ in the large--$N$ limit,
generated by the (non--zero) coupling constant $c_1$.

In section 3, generalizing the results obtained in Refs. \cite{EM2003,EM2004},
where the effects of the new $U(1)$ chiral condensate on the radiative decays
of the pseudoscalar mesons $\eta$ and $\eta'$ into two photons had been
investigated, we have studied the effects of the $U(1)$ chiral condensate
on the strong decays of the ``light'' pseudoscalar mesons, i.e.,
$\eta,\eta' \to 3\pi^0$; $\eta,\eta' \to \pi^+ \pi^- \pi^0$;
$\eta' \to \eta \pi^0 \pi^0$; $\eta' \to \eta \pi^+ \pi^-$;
and also on the strong decays of the exotic (``heavy'') $SU(3)$--singlet
pseudoscalar state $\eta_X$:
$\eta_X \to 3\pi^0$; $\eta_X \to \pi^+ \pi^- \pi^0$;
$\eta_X \to \eta \pi^0 \pi^0$; $\eta_X \to \eta \pi^+ \pi^-$;
$\eta_X \to \eta' \pi^0 \pi^0$; $\eta_X \to \eta' \pi^+ \pi^-$;
$\eta_X \to 3\eta, 3 \eta', \eta\eta\eta', \eta\eta'\eta'$.
Concerning the decays of the exotic particle $\eta_X$, we have found
some relations between its mass and its decay widths, which in principle
might be useful to identify a possible candidate for this particle.
According to the \emph{Particle Data Group} \cite{PDG2010}, the possible
candidates for the $\eta_X$, having the same quantum numbers
$I^G(J^{PC})=0^+(0^{-+})$ of the $\eta'$, but a larger mass, are,
at the moment, those reported in Eq. \eqref{candidates}
(other candidates with larger masses are also
present, but some of their quantum numbers $I^G(J^{PC})$ are not yet known):
unfortunately, no quantitative determination of their decay widths into (e.g.)
three pions has been done up to now.

Concerning the decays $\eta,\eta' \rightarrow 3\pi$, it is well known that,
because of large unitarity corrections due to strong final--state interactions,
one has to go beyond leading and even one--loop order in chiral perturbation
theory in order to obtain a valid, reliable representation of the
decay amplitudes and of the corresponding decay widths, that
can be successfully compared with the experimental values
\cite{GL1985,KWW1996,AL1996,BB2003,BN2005,BG2007}.\\
In the present paper we have not, of course, aimed at that. In particular,
we could not proceed as in the case of the radiative decays
$\eta,\eta'\rightarrow \gamma\gamma$, i.e., we could not extract the value of
$\fx$ (and of the mixing angle $\tilde\varphi$) by comparing, e.g., the
leading--order theoretical predictions \eqref{gamma0-eta} and
\eqref{gamma0-eta'}, for the $\eta \rightarrow 3\pi^0$ and
$\eta' \rightarrow 3\pi^0$ decay widths, with the corresponding experimental
values reported in Table 1 of section 3.\\
Instead, our aim has been simply to quantify the corrections coming from the
non--zero value $\fx=24(7)$ MeV, that we have found studying the radiative
decays $\eta,\eta'\rightarrow\gamma\gamma$ [see the Introduction and in
particular Eq. \eqref{results}], taking the leading--order amplitudes/widths
in the $\fx=0$ case as a useful reference point.
From the values reported in Table 1 of section 3 we have concluded that:
\begin{itemize}
\item[i)] In the case of the $\eta \rightarrow 3\pi$ decays, the size of
the corrections $\Delta\Gamma_{\rm LO}$ coming from a non--zero value $\fx =
24(7)$ MeV, with respect to the $\fx=0$ case, is very small, being of the
order of $1\%$, i.e., comparable to (or even smaller than) the size of the
{\it electromagnetic} corrections for these decays, which have been recently
re--calculated in Ref. \cite{DKM2009}.
\item[ii)] Instead, in the case of the $\eta' \rightarrow 3\pi$ decays, the
size of the corrections $\Delta\Gamma_{\rm LO}$ is much larger, being of the
order of $30\%$. Moreover, at least for the decay $\eta' \rightarrow 3\pi^0$
(where the statistical errors are smaller), this (negative) correction seems
to go in the right direction, improving the agreement between the theoretical
prediction and the experimental value.
\end{itemize}

Finally, concerning the decays $\eta' \rightarrow \eta \pi^0 \pi^0$ and
$\eta' \rightarrow \eta \pi^+ \pi^-$, knowing already from Ref. \cite{DNPV1981}
that the leading--order theoretical predictions for $\fx=0$
are in strong disagreement with the experimental values,
we have tried to see if the introduction of a non--zero value of $\fx$ can cure,
at least in part, this disagreement: but we have found that it cannot. In fact,
even if the correction $\Delta\Gamma_{\rm LO}$ is quite large (of the order
of $30\%$) if compared with the value of $\Gamma_{\rm LO}$ at $\fx=0$,
it is, however, too small if compared with the experimental value, and,
moreover, being negative, it goes in the ``wrong'' direction, lowering the
theoretical prediction at $\fx=0$, which is already much smaller than the
experimental value. (In other words, it is not possible to find a value of
the parameter $\fx$ which moves the leading--order theoretical prediction
towards the experimental value.)

However, as we have already stressed in the conclusions of Refs.
\cite{EM2003,EM2004}, one should keep in mind that our results have been
derived from a very simplified model, obtained by doing a first--order
expansion in $1/N$ and in the quark masses. We expect that such a model can
furnish only qualitative or, at most, ``semi--quantitative'' predictions.
As already observed in Ref. \cite{DNPV1981},
in order to obtain a better agreement with the experimental data of the decay
widths, most probably it is not enough to retain only the leading order in the
$1/N$ expansion, but one has to go to the next--to--leading order.
The introduction, in our model, of higher--order terms
in the $1/N$ expansion is, of course, a quite hard task, which is beyond the
aim of the present paper. Further studies are therefore necessary in order
to continue this analysis from a more quantitative point of view.
We expect that some progress will be made along this line in the near future.

\section*{Acknowledgements}

The author is extremely grateful to D. Somensi for his help during the
initial stage of the work.

\newpage

\renewcommand{\thesection}{}
\renewcommand{\thesubsection}{A.\arabic{subsection} }
 
\pagebreak[3]
\setcounter{section}{1}
\setcounter{equation}{0}
\setcounter{subsection}{0}
\setcounter{footnote}{0}

\begin{flushleft}
{\Large\bf \thesection Appendix A: The $U(1)$ chiral order parameter}
\end{flushleft}

\renewcommand{\thesection}{A}

\noindent
We make the assumption (discussed in the Introduction)
that the $U(1)$ chiral symmetry is broken independently from the
$SU(L) \otimes SU(L)$ symmetry. The usual chiral order parameter
$\langle \bar{q} q \rangle$ is an order parameter both for
$SU(L) \otimes SU(L)$ and for $U(1)_A$: when it is different from zero,
$SU(L) \otimes SU(L)$ is broken down to $SU(L)_V$ (``$V$'' stands for
{\it ``vectorial''}) and also $U(1)_A$ is broken.
In fact, under a $U(1)$ chiral transformation with parameter $\alpha$
(as usual, $q_L \equiv \frac{1}{2}(1 + \gamma_5)q$ and
$q_R \equiv \frac{1}{2}(1 - \gamma_5)q$, with
$\gamma_5 \equiv -i \gamma^0 \gamma^1 \gamma^2 \gamma^3$, denote respectively
the {\it left--handed} and the {\it right--handed} quark fields):
\be
U(1)_A: \quad q \to e^{-i\alpha\gamma_5}q, \quad {\rm i.e.}, \quad
q_L \to e^{-i\alpha}q_L, \quad q_R \to e^{i\alpha}q_R ,
\label{trasfu1}
\ee
the chiral condensate would transform as (assuming the $U(1)_A$ symmetry to
be realized {\it \`a la} Wigner--Weyl):
\be
U(1)_A: \quad \langle \bar{q} q\rangle \rightarrow e^{2i \alpha}
\langle \bar{q}_L q_R\rangle + e^{-2i \alpha}\langle \bar{q}_R q_L\rangle.
\ee
By taking $\alpha = \pi/2$, we would obtain
$\langle \bar{q} q\rangle \rightarrow -\langle \bar{q} q\rangle$:
therefore, if the chiral condensate is different
from zero, the $U(1)_A$ symmetry cannot be realized {\it \`a la} Wigner--Weyl.
Thus we need another quantity which could be an order parameter only for 
the $U(1)$ chiral symmetry \cite{EM1994a,EM1994b,EM1994c,EM1995}.
The most simple quantity of this kind was introduced by Kobayashi and Maskawa
in 1970 \cite{KM1970}, as an additional effective vertex in a generalized
Nambu--Jona--Lasinio model, and it was later derived by 'tHooft in 1976
\cite{tHooft1976}, as an instanton--induced quark interaction.
(See also Ref. \cite{Kunihiro2009} for an historical review on this subject.)\\
For a theory with $L$ light quark flavours (of mass $m_i \ll \Lambda_{QCD}$;
$i=1,\ldots ,L$), it is a $2L$--fermion interaction that has the chiral 
transformation properties of:
\be
{\cal O}_{U(1)}^{(L)} \sim \displaystyle {\det_{st}} \left[ \bar{q}_s
\left( \frac{1+\gamma_5}{2} \right) q_t \right] + {\rm H.c.} =
\displaystyle {\det_{st}} \left( \bar{q}_{sR} q_{tL} \right) +
\displaystyle {\det_{st}} \left( \bar{q}_{sL} q_{tR} \right) ,
\ee
where $s,t = 1, \ldots ,L$ are flavour indices, but the colour indices are 
arranged in a more general way (see below).
Since under chiral $U(L) \otimes U(L)$ transformations the quark fields
transform as follows:
\be
U(L) \otimes U(L):~~~ q_L \to V_L q_L ~~~,~~~q_R \to V_R q_R,
\ee
where $V_L$ and $V_R$ are arbitrary $L \times L$ unitary matrices, 
we immediately derive the transformation property of
${\cal O}_{U(1)}^{(L)}$ under $U(L) \otimes U(L)$:
\be
U(L) \otimes U(L):~~~ {\cal O}_{U(1)}^{(L)} \to \det(V_L)\det(V_R)^*
\displaystyle {\det_{st}} \left( \bar{q}_{sR} q_{tL} \right) + {\rm H.c.}
\ee
This just means that ${\cal O}_{U(1)}^{(L)}$ is invariant under
$SU(L) \otimes SU(L) \otimes U(1)_V$, while it is not invariant under the
$U(1)_A$ transformation \eqref{trasfu1}:
\be
U(1)_A:~~~ {\cal O}_{U(1)}^{(L)} \to e^{-i 2L \alpha}
\displaystyle {\det_{st}} \left( \bar{q}_{sR} q_{tL} \right) + {\rm H.c.}
\ee

\subsection{The $U(1)$ chiral condensate for $L=2$ }

\noindent
As an example let us consider the most simple case, that is $L=2$, but with 
a general colour group $SU(N)$.
It is not hard to find (using the Fierz relations both for the spinorial 
matrices and the $SU(N)$ generators in their fundamental representation) 
that the most general colour--singlet, Hermitian and $P$--invariant 
{\it local} quantity (without derivatives)
which has the required chiral transformation properties is just the following
four--fermion local operator:
\be
{\cal O}_{U(1)}^{(L=2)}(\alpha_0 ,\beta_0) = 
F_{bd}^{ac}(\alpha_0 ,\beta_0) \epsilon^{st}
\left( \bar{q}_{1R}^{a}q_{sL}^{b} \cdot
\bar{q}_{2R}^{c}q_{tL}^{d} +
\bar{q}_{1L}^{a}q_{sR}^{b} \cdot
\bar{q}_{2L}^{c}q_{tR}^{d} \right) ,
\label{opl2}
\ee
where the colour tensor $F_{bd}^{ac}(\alpha_0 ,\beta_0)$ is given by:
\be
F_{bd}^{ac}(\alpha_0 ,\beta_0) = 
\alpha_0 \delta_b^a \delta_d^c + \beta_0 \delta_d^a \delta_b^c,
\label{Facbd}
\ee
$\alpha_0$ and $\beta_0$ being arbitrary real parameters. 
In Eq. (\ref{opl2}), $a,b,c,d \in \{1, \ldots ,N\}$ are colour indices;
$s,t \in \{1,2\}$ are flavour indices and $\epsilon^{st}=-\epsilon^{ts}$,
$\epsilon^{12}=1$. Dirac indices are contracted between the first and the 
second fermion field and also between the third and the fourth one.
Note that if we choose $\alpha_0 = N$ and $\beta_0 = -1$, 
${\cal O}_{U(1)}^{(L=2)}(\alpha_0 ,\beta_0)$ 
just becomes (up to a proportionality constant) the effective Lagrangian
for two flavours of quarks in an instanton background, found by 'tHooft
in \cite{tHooft1976} .

Now, to obtain an order parameter for the $U(1)$ chiral symmetry, one can 
simply take the vacuum expectation value of 
${\cal O}_{U1)}^{(L=2)}(\alpha_0 ,\beta_0)$:
\be
C_{U(1)}^{(L=2)}(\alpha_0 ,\beta_0) \equiv 
\langle {\cal O}_{U(1)}^{(L=2)}(\alpha_0 ,\beta_0) \rangle .
\label{cl2}
\ee
The arbitrarity in the choice of $\alpha_0$ and $\beta_0$ (indeed of only 
one of them, since only their ratio is relevant) can be removed if we 
require that the new $U(1)$ chiral condensate is ``independent'', in a 
sense which will be explained below, of the usual chiral condensate
$\langle \bar{q} q \rangle$.
As it was pointed out by Shifman, Vainshtein and Zakharov in 
\cite{SVZ1979}, a matrix element of the form 
$\langle \bar{q} \Gamma_1 q \cdot \bar{q} \Gamma_2 q \rangle$ 
has, in general, a contribution proportional to the square 
of the vacuum expectation value (v.e.v.) of $\bar{q} q$. This 
contribution corresponds to retaining the vacuum intermediate state in all 
the channels and neglecting the contributions of all the other states; we 
call this contribution the {\it ``disconnected part''} of the original matrix 
element:
\be
\langle \bar{q} \Gamma_1 q \cdot \bar{q} \Gamma_2 q \rangle_{disc} 
= \frac{1}{G^2} \left[ (\Tr\Gamma_1 \cdot \Tr\Gamma_2) - 
\Tr(\Gamma_1 \Gamma_2) \right] \langle \bar{q} q \rangle^2,
\ee
where the normalization factor $G$ is defined as
($\bar{q} q = \sum_A \bar{q}_A q_A$):
\be
\langle \bar{q}_A q_B \rangle =
\frac{\delta_{AB}}{G} \langle \bar{q} q \rangle, \quad {\rm i.e.}, \quad
G = \delta_{AA},
\label{norm1}
\ee
and the subscripts $A,B$ are collective indices which include spin, colour and
flavour; therefore, $G = 4 \times L \times N$ for a general $L$, and $G=8N$
for $L=2$.
When considering the operator ${\cal O}_{U(1)}^{(L=2)}(\alpha_0 ,\beta_0)$ 
defined in Eqs. (\ref{opl2}) and (\ref{Facbd}), we find the following
expression for its {\it disconnected} part:
\be
\langle {\cal O}^{(L=2)}_{U(1)}(\alpha_0 ,\beta_0) \rangle_{disc} =
\frac{1}{16N} [N (2\alpha_0 + \beta_0) + (\alpha_0 + 2\beta_0)]
\langle \bar{q} q \rangle^2 ,
\ee
where: $\langle \bar{q} q \rangle = \langle \bar{u} u \rangle
+ \langle \bar{d} d \rangle$.
From this last equation we immediately see that the {\it disconnected}
part of the condensate $C_{U(1)}^{(L=2)}(\alpha_0 ,\beta_0)$ vanishes
with the following particular choice of the coefficients $\alpha_0$ and
$\beta_0$ (only their ratio is really relevant):
\be
\frac{\beta_0}{\alpha_0} = -\frac{2N + 1}{N + 2}.
\label{alphabeta}
\ee 
In other words, the condensate (\ref{cl2}) with $\alpha_0$ and $\beta_0$
satisfying the constraint (\ref{alphabeta}) does not take contributions
from the usual chiral condensate $\langle \bar{q} q \rangle$.
To summarize, a good choice for a $U(1)$ chiral condensate which is really
``independent'' of the usual chiral condensate $\langle \bar{q} q \rangle$
is the following one (apart from an irrelevant multiplicative constant):
\be
C_{U(1)}^{(L=2)} = 
\langle ( \delta_b^a \delta_d^c - \frac{2N + 1}{N + 2}
\delta_d^a \delta_b^c ) 
\epsilon^{st}
\left( \bar{q}_{1R}^{a}q_{sL}^{b} \cdot
\bar{q}_{2R}^{c}q_{tL}^{d} +
\bar{q}_{1L}^{a}q_{sR}^{b} \cdot
\bar{q}_{2L}^{c}q_{tR}^{d} \right) \rangle .
\label{condl2}
\ee
As a remark, we observe that the condensate $C_{U(1)}^{(L=2)}$ so defined
turns out to be of order ${\cal O}(g^2 N^2) = {\cal O}(N)$ in the
large--$N$ expansion (this result was also derived in Ref. \cite{EM1994b} by
simply requiring that the $1/N$ expansion of the relevant QCD Ward Identities
remains well defined when including this new condensate).
In the case of physical interest, i.e., $N =3$, the condensate
Eq. (\ref{condl2}) becomes:
\be
C_{U(1)}^{(L=2)} = 
\langle ( \delta_b^a \delta_d^c - \frac{7}{5} \delta_d^a \delta_b^c )
\epsilon^{st} \left( \bar{q}_{1R}^{a}q_{sL}^{b} \cdot
\bar{q}_{2R}^{c}q_{tL}^{d} + \bar{q}_{1L}^{a}q_{sR}^{b} \cdot
\bar{q}_{2L}^{c}q_{tR}^{d} \right) \rangle .
\ee

\subsection{The $U(1)$ chiral condensate for $L=3$} 

\noindent
So far we have considered the most simple case $L=2$. However, this
procedure can be easily generalized to every $L$, and we can take as
an order parameter for the $U(1)$ chiral symmetry:
\be
C^{(L)}_{U(1)} = \langle {\cal O}_{U(1)}^{(L)} \rangle .
\ee
As we have done in the case $L=2$, the colour indices may be arranged in such
a way that the $U(1)$ chiral condensate does not take contributions from the
usual chiral condensate $\langle \bar{q} q \rangle$: as a consenquence of
this, the new condensate will be of order ${\cal O}(g^{2L - 2} N^L) =
{\cal O}(N)$ in the large--$N$ expansion \cite{EM1994b,EM1994c}.\\
In the {\it real--world} case there are $L=3$ light flavours, $u$, $d$, and
$s$, with masses $m_u$, $m_d$, and $m_s$ which are small compared to the QCD
mass--scale $\Lambda_{QCD}$.
Proceeding as in the case $L=2$ [see Eq. \eqref{opl2}], one reduces to
consider the following general colour--singlet, Hermitian and $P$--invariant 
{\it local} six--fermion operator (without derivatives):
\be
{\cal O}_{U(1)}^{(L=3)} = F^{a_1 a_2 a_3}_{b_1 b_2 b_3} \epsilon^{l_1 l_2 l_3}
\ \bar{q}^{\ a_1}_{1R} \ q^{b_1}_{l_1 L} \cdot
\bar{q}^{\ a_2}_{2R} \ q^{b_2}_{l_2 L} \cdot
\bar{q}^{\ a_3}_{3R} \ q^{b_3}_{l_3 L} + {\rm H.c.},
\label{O6}
\ee
where $a_1,a_2,a_3,b_1,b_2,b_3 \in \{1,2, \cdots N\}$ are colour indices,
$l_1,l_2,l_3 \in \{1,2,3\}$ are \emph{flavour} indices and the colour tensor
$F^{a_1 a_2 a_3}_{b_1 b_2 b_3}$ is given by:
\bea
F^{a_1 a_2 a_3}_{b_1 b_2 b_3} &=& 
\alpha_1 \delta^{a_1}_{b_1}\delta^{a_2}_{b_2} \delta^{a_3}_{b_3} +
\alpha_2 \delta^{a_1}_{b_2}\delta^{a_2}_{b_3} \delta^{a_3}_{b_1} +
\alpha_3 \delta^{a_1}_{b_3}\delta^{a_2}_{b_1} \delta^{a_3}_{b_2} \nonumber \\
&+& \beta_1 \delta^{a_1}_{b_2}\delta^{a_2}_{b_1} \delta^{a_3}_{b_3} +
\beta_2 \delta^{a_1}_{b_1}\delta^{a_2}_{b_3} \delta^{a_3}_{b_2} +
\beta_3 \delta^{a_1}_{b_3}\delta^{a_2}_{b_2} \delta^{a_3}_{b_1} ,
\label{F}
\eea
with $\alpha_1,\alpha_2,\alpha_3,\beta_1,\beta_2,\beta_3$ real
parameters. However, differently from the case $L=2$, the operator
${\cal O}_{U(1)}^{(L=3)}$ in Eqs. \eqref{O6}--\eqref{F}, with {\it arbitrary}
real parameters $\alpha_1,\alpha_2,\alpha_3,\beta_1,\beta_2,\beta_3$, is
{\it not}, in general, invariant under a $SU(3) \otimes SU(3)$ chiral
transformation:
\be
SU(3) \otimes SU(3):~~~ q_L \to U_L q_L ~~~,~~~q_R \to U_R q_R
\ee
$(\det U_L = \det U_R = 1)$. Invariance under $SU(3) \otimes SU(3)$ is,
instead, recovered provided that the colour tensor
$F^{a_1 a_2 a_3}_{b_1 b_2 b_3}$ satisfies the following {\it symmetry property}:
\be
F^{a_1 a_2 a_3}_{b_1 b_2 b_3} = F^{a_i a_j a_k}_{b_i b_j b_k},
\quad \forall \text{ permutations } \{i,j,k\} \text{ of } \{1,2,3\} .
\label{fprop}
\ee
In fact, in this case it is easy to see that the operator \eqref{O6} can
be re--written in the following form:
\be
{\cal O}_{U(1)}^{(L=3)} = F^{a_1 a_2 a_3}_{b_1 b_2 b_3}
\frac{1}{3!} \epsilon^{r_1 r_2 r_3} \epsilon^{l_1 l_2 l_3} \
\bar{q}^{\ a_1}_{r_1 R} \ q^{b_1}_{{l_1 L}} \cdot
\bar{q}^{\ a_2}_{r_2 R} \ q^{b_2}_{{l_2 L}} \cdot
\bar{q}^{\ a_3}_{r_3 R} \ q^{b_3}_{{l_3 L}} + {\rm H.c.},
\label{opl3}
\ee
which is manifestly invariant under $SU(3) \otimes SU(3)$:
\bea
{\cal O}_{U(1)}^{(L=3)} &\rightarrow&
F^{a_1 a_2 a_3}_{b_1 b_2 b_3}
\frac{1}{3!} \epsilon^{r_1 r_2 r_3} \epsilon^{l_1 l_2 l_3}
\bar{q}^{\ a_1}_{s_1 R} (U_R^{\dagger})_{s_1 r_1} (U_L)_{l_1 m_1} \
q^{b_1}_{m_1 L} \nonumber \\
&\cdot& \bar{q}^{\ a_2}_{s_2 R} (U^{\dagger}_R)_{s_2 r_2} (U_L)_{l_2 m_2} \
q^{b_2}_{m_2 L} \cdot \bar{q}^{\ a_3}_{s_3 R} (U^{\dagger})_{s_3 r_3}
(U_L)_{l_3 m_3} \ q^{b_3}_{m_3 L} + {\rm H.c.} \nonumber \\
&=& F^{a_1 a_2 a_3}_{b_1 b_2 b_3} \frac{1}{3!}
\det(U_R^\dagger) \epsilon^{s_1 s_2 s_3}
\det(U_L) \epsilon^{m_1 m_2 m_3} \nonumber \\
&\times& \bar{q}^{\ a_1}_{s_1 R} \ q^{b_1}_{{m_1 L}} \cdot
\bar{q}^{\ a_2}_{s_2 R} \ q^{b_2}_{{m_2 L}} \cdot
\bar{q}^{\ a_3}_{s_3 R} \ q^{b_3}_{{m_3 L}} + {\rm H.c.} \nonumber \\
&=& F^{a_1 a_2 a_3}_{b_1 b_2 b_3} \frac{1}{3!}
\epsilon^{s_1 s_2 s_3} \epsilon^{m_1 m_2 m_3}
\bar{q}^{\ a_1}_{s_1 R} \ q^{b_1}_{{m_1 L}} \cdot
\bar{q}^{\ a_2}_{s_2 R} \ q^{b_2}_{{m_2 L}} \cdot
\bar{q}^{\ a_3}_{s_3 R} \ q^{b_3}_{{m_3 L}} + {\rm H.c.} \nonumber \\
&=& {\cal O}_{U(1)}^{(L=3)} .
\label{su3inv}
\eea
(Or, equivalently, one can start from the expression \eqref{opl3} of the
six--fermion local operator, which, on the basis of \eqref{su3inv}, is
invariant under $SU(3) \otimes SU(3)$ for {\it every} choice of the colour
tensor $F^{a_1 a_2 a_3}_{b_1 b_2 b_3}$: but one immediately recognizes that
only the {\it symmetric} part of the colour tensor, satisfying the relation
\eqref{fprop}, contributes to the r.h.s. of \eqref{opl3}, the
{\it anti--symmetric} parts being trivially cancelled out.
Note that, in the case $L=2$, the most general colour tensor \eqref{Facbd}
automatically satifies the {\it symmetry property},
$F^{ac}_{bd} = F^{ca}_{db}$.)\\
The symmetry property \eqref{fprop} imposes the following constraints on
the parameters of the colour tensor \eqref{F}:
\be
\alpha_3 = \alpha_2, \qquad \beta_3 = \beta_2 = \beta_1 .
\label{vincoli1}
\ee
The colour tensor has, therefore, the following form:
\bea
F^{a_1 a_2 a_3}_{b_1 b_2 b_3} &=& 
\alpha_1 \delta^{a_1}_{b_1} \delta^{a_2}_{b_2} \delta^{a_3}_{b_3} +
\alpha_2 (\delta^{a_1}_{b_2} \delta^{a_2}_{b_3} \delta^{a_3}_{b_1} +
\delta^{a_1}_{b_3} \delta^{a_2}_{b_1} \delta^{a_3}_{b_2}) \nonumber \\
&+& \beta_1 (\delta^{a_1}_{b_2} \delta^{a_2}_{b_1} \delta^{a_3}_{b_3} +
\delta^{a_1}_{b_1} \delta^{a_2}_{b_3} \delta^{a_3}_{b_2} +
\delta^{a_1}_{b_3} \delta^{a_2}_{b_2} \delta^{a_3}_{b_1}) ,
\label{F3}
\eea
in terms of three arbitrary real parameters $\alpha_1$, $\alpha_2$, $\beta_1$.\\
Let us now evaluate the vacuum expectation value of the operator
${\cal O}_{U(1)}^{(L=3)}$:
\bea
C^{(L=3)}_{U(1)} &\equiv& \langle {\cal O}_{U(1)}^{(L=3)} \rangle =
F^{a_1 a_2 a_3}_{b_1 b_2 b_3} \epsilon^{l_1 l_2 l_3}
\langle \bar{q} \Gamma_1 q \cdot \bar{q} \Gamma_2 q \cdot
\bar{q} \Gamma_3 q \rangle + {\rm c.c.} \nonumber \\ 
&=& F^{a_1 a_2 a_3}_{b_1 b_2 b_3} \epsilon^{l_1 l_2 l_3}
(\Gamma_1)_{AB} (\Gamma_2)_{CD} (\Gamma_3)_{EF}
\langle \bar{q}_A q_B \bar{q}_C q_D \bar{q}_E q_F \rangle
+ {\rm c.c.},
\label{C3}
\eea
where [see Eq. \eqref{O6}]:
\bea
(\Gamma_1)_{AB} = (\Gamma_1)_{i_1 j_1, m_1 n_1}^{c_1 d_1} &=&
\left( \frac{1 + \gamma_5}{2} \right)_{i_1 j_1}
\otimes (\delta_{m_1 1}\delta_{n_1 l_1}) \otimes (\delta^{c_1 a_1}
\delta^{d_1 b_1}), \nonumber \\
(\Gamma_2)_{CD} = (\Gamma_2)_{i_2 j_2, m_2 n_2}^{c_2 d_2} &=&
\left( \frac{1 + \gamma_5}{2} \right)_{i_2 j_2}
\otimes (\delta_{m_2 2}\delta_{n_2 l_2}) \otimes (\delta^{c_2 a_2}
\delta^{d_2 b_2}), \nonumber \\
(\Gamma_3)_{EF} = (\Gamma_3)_{i_3 j_3, m_3 n_3}^{c_3 d_3} &=&
\left( \frac{1 + \gamma_5}{2} \right)_{i_3 j_3}
\otimes (\delta_{m_3 3}\delta_{n_3 l_3}) \otimes (\delta^{c_3 a_3}
\delta^{d_3 b_3}),
\label{gamma123}
\eea
where $i,j$ are Dirac indices, $m,n$ are flavour indices, and $c,d$ are
colour indices.\\
As in the case $L=2$ treated above, we can write the vacuum expectation value
of the operator ${\cal O}_{U(1)}^{(L=3)}$ as the sum of a
\emph{connected part}, which does \emph{not} depend on the chiral condensate
$\langle \bar{q}q \rangle$, and a \emph{disconnected part}, which instead
contains the chiral condensate $\langle \bar{q}q \rangle$, i.e.,
$C_{U(1)}^{(L=3)} = \langle {\cal O}_{U(1)}^{(L=3)} \rangle _{conn} +
\langle {\cal O}_{U(1)}^{(L=3)} \rangle _{disc}$, where:
\be
\langle {\cal O}_{U(1)}^{(L=3)} \rangle _{disc} =
F^{a_1 a_2 a_3}_{b_1 b_2 b_3}\epsilon^{l_1 l_2 l_3}
(\Gamma_1)_{AB} (\Gamma_2)_{CD} (\Gamma_3)_{EF}
\langle \bar{q}_A q_B \bar{q}_C q_D \bar{q}_E q_F \rangle_{disc} + {\rm c.c.},
\ee
and the \emph{disconnected part} of the v.e.v. of the six--fermion operator
has the following form:
\bea
\lefteqn{
\langle \bar{q}_A q_B \bar{q}_C q_D \bar{q}_E q_F \rangle_{disc} }
\nonumber \\
& & = \langle \bar{q}_A q_B \rangle
\langle \bar{q}_C q_D \bar{q}_E q_F \rangle_{conn} +
\langle \bar{q}_C q_D \rangle
\langle \bar{q}_A q_B \bar{q}_E q_F \rangle_{conn} +
\langle \bar{q}_E q_F \rangle
\langle \bar{q}_A q_B \bar{q}_C q_D \rangle_{conn} \nonumber \\
& & - \langle \bar{q}_A q_D \rangle
\langle \bar{q}_C q_B \bar{q}_E q_F \rangle_{conn} - 
\langle \bar{q}_A q_F \rangle
\langle \bar{q}_C q_D \bar{q}_E q_B \rangle_{conn} -
\langle \bar{q}_C q_B \rangle
\langle \bar{q}_A q_D \bar{q}_E q_F \rangle_{conn} \nonumber\\
& & - \langle \bar{q}_C q_F \rangle
\langle \bar{q}_A q_B \bar{q}_E q_D \rangle_{conn} -
\langle \bar{q}_E q_B \rangle
\langle \bar{q}_A q_F \bar{q}_C q_D \rangle_{conn} -
\langle \bar{q}_E q_D \rangle
\langle \bar{q}_A q_B \bar{q}_C q_F \rangle_{conn} \nonumber\\
& & + \langle \bar{q}_A q_B \rangle \langle \bar{q}_C q_D \rangle
\langle \bar{q}_E q_F \rangle -
\langle \bar{q}_A q_B \rangle \langle \bar{q}_C q_F \rangle
\langle \bar{q}_E q_D \rangle -
\langle \bar{q}_A q_D \rangle \langle \bar{q}_C q_B \rangle
\langle \bar{q}_E q_F \rangle \nonumber\\
& & + \langle \bar{q}_A q_D \rangle \langle \bar{q}_C q_F \rangle
\langle \bar{q}_E q_B \rangle +
\langle \bar{q}_A q_F \rangle \langle \bar{q}_C q_B \rangle
\langle \bar{q}_E q_D \rangle -
\langle \bar{q}_A q_F \rangle \langle \bar{q}_C q_D \rangle
\langle \bar{q}_E q_B \rangle .
\label{discon}
\eea
On the basis of Eq. \eqref{norm1}, we see that the disconnected part of the
condensate \eqref{C3} can be written as:
\be
\langle {\cal O}_{U(1)}^{(L=3)} \rangle _{disc} =
A_1 \langle \bar{q} q \rangle + A_3 \langle \bar{q} q \rangle ^3 ,
\label{partedisc}
\ee
where the first term ($A_1 \langle \bar{q} q \rangle$), proportional to
the chiral condensate, is originated by the first nine terms in the r.h.s.
of Eq. \eqref{discon}, while the second term
($A_3\langle \bar{q} q \rangle ^3$) is originated by the last six terms
in the r.h.s. of Eq. \eqref{discon} and represents the \emph{completely
disconnected} part, proportional to the third power of the chiral condensate.\\
Explicitly, using Eq. \eqref{norm1}, with $G=4 \times 3 \times N=12N$, the
form \eqref{gamma123} of the $\Gamma$ matrices and the form \eqref{F3} of
the colour tensor $F^{a_1 a_2 a_3}_{b_1 b_2 b_3}$, satisfying the symmetry
property \eqref{fprop}, we obtain the following expression for the
coefficient $A_1$:
\bea
A_1 &=& \frac{1}{G^3} F^{a_1 a_2 a_3}_{b_1 b_2 b_3} \epsilon^{l_1 l_2 l_3}
\{ \Tr \Gamma_1 \langle \bar{q} \Gamma_2 q \cdot \bar{q} \Gamma_3 q
\rangle_{conn} \nonumber \\
&+& \Tr \Gamma_2 \langle \bar{q} \Gamma_1 q \cdot \bar{q} \Gamma_3 q
\rangle_{conn}
+ \Tr \Gamma_3 \langle \bar{q} \Gamma_1 q \cdot \bar{q} \Gamma_2 q
\rangle_{conn} \nonumber \\
&-& \langle \bar{q} \Gamma_1 \Gamma_2 q \cdot \bar{q} \Gamma_3 q
\rangle_{conn}
- \langle \bar{q} \Gamma_2 \Gamma_1 q \cdot \bar{q} \Gamma_3 q
\rangle_{conn} \nonumber \\
&-& \langle \bar{q} \Gamma_1 \Gamma_3 q \cdot \bar{q} \Gamma_2 q
\rangle_{conn}
- \langle \bar{q} \Gamma_3 \Gamma_1 q \cdot \bar{q} \Gamma_2 q
\rangle_{conn} \nonumber \\
&-& \langle \bar{q} \Gamma_2 \Gamma_3 q \cdot \bar{q} \Gamma_1 q
\rangle_{conn}
- \langle \bar{q} \Gamma_3 \Gamma_2 q \cdot \bar{q} \Gamma_1 q
\rangle_{conn} \} + {\rm c.c.} \nonumber \\
&=& \frac{1}{12N} \left[ 2 F^{ace}_{bde} + F^{ace}_{bed} + F^{ace}_{edb}
\right] \left( C_{12}^{abcd} + C_{13}^{abcd} + C_{23}^{abcd} \right) ,
\label{A1}
\eea
where:
\bea
C_{12}^{abcd} &\equiv& \epsilon^{st3} \left[ \langle
\bar{q}^{\ a}_{1R} \ q^{b}_{sL} \cdot
\bar{q}^{\ c}_{2R} \ q^{d}_{tL} \rangle_{conn} + {\rm c.c.} \right] ,
\nonumber \\
C_{13}^{abcd} &\equiv& \epsilon^{s2t} \left[ \langle
\bar{q}^{\ a}_{1R} \ q^{b}_{sL} \cdot
\bar{q}^{\ c}_{3R} \ q^{d}_{tL} \rangle_{conn} + {\rm c.c.} \right] ,
\nonumber \\
C_{23}^{abcd} &\equiv& \epsilon^{1st} \left[ \langle
\bar{q}^{\ a}_{2R} \ q^{b}_{sL} \cdot
\bar{q}^{\ c}_{3R} \ q^{d}_{tL} \rangle_{conn} + {\rm c.c.} \right] ,
\label{C123}
\eea
and the following expression for the coefficient $A_3$:
\bea
A_3 &=& \frac{2}{G^3} F^{a_1 a_2 a_3}_{b_1 b_2 b_3} \epsilon^{l_1 l_2 l_3}
\left[ \Tr \Gamma_1 \Tr \Gamma_2 \Tr \Gamma_3
- \Tr \Gamma_1 \Tr (\Gamma_2 \Gamma_3)
- \Tr (\Gamma_1 \Gamma_2) \Tr \Gamma_3 \right. \nonumber \\
&+& \left. \Tr (\Gamma_1 \Gamma_3 \Gamma_2)
+ \Tr (\Gamma_1 \Gamma_2 \Gamma_3) - \Tr (\Gamma_1 \Gamma_3) \Tr \Gamma_2
\right] \nonumber \\
&=& \frac{1}{216 N^3}[\alpha_1 (2N^3 + 3N^2 + N)
+ \alpha_2 (N^3 + 6N^2 + 5N) \nonumber \\
&+& \beta_1 (3N^3 + 9N^2 + 6N) ].
\label{A3}
\eea
Now, if we want to obtain a new order parameter which is really
\emph{independent} on the usual chiral condensate $\langle \bar{q}q \rangle$,
we must require that its disconnected part \eqref{partedisc} vanishes
independently on the value of $\langle \bar{q}q \rangle$, imposing the two
conditions $A_1=0$ and $A_3=0$. Therefore, we have two independent constraints
on the three parameters $\alpha_1$, $\alpha_2$ and $\beta_1$, which enter the
colour tensor \eqref{F3}: the new condensate $C_{U(1)}^{(L=3)}$ is then
univocally determined, apart from a multiplicative constant.\\
Let us also observe that in the large--$N$ limit, taking the coefficients
$\alpha_1$, $\alpha_2$ and $\beta_1$ in the colour tensor \eqref{F3} to
be of order ${\cal O}(N^0)$, the coefficient $A_1$ is of order ${\cal O}(N)$,
while the coefficient $A_3$ is of order ${\cal O}(N^0)$: and, consequently,
the first term $A_1 \langle \bar{q} q \rangle$ in the r.h.s. of
\eqref{partedisc} is of order ${\cal O}(N^2)$, while the second term
$A_3\langle \bar{q} q \rangle ^3$ is of order ${\cal O}(N^3)$
(being $\langle \bar{q}q \rangle = \OO(N)$). If both these disconnected parts
are zero, then the new condensate $C_{U(1)}^{(L=3)}$ is simply equal to the
\emph{connected} part $\langle {\cal O}_{U(1)}^{(L=3)} \rangle _{conn}$,
which is of order ${\cal O}(N)$, i.e., of the same order of the usual
chiral condensate $\langle \bar{q}q \rangle$ (as already observed in
Refs. \cite{EM1994b,EM1994c}).\\
We also observe that the condition $A_1=0$ implies that the new six--fermion
condensate $C_{U(1)}^{(L=3)}$ does not take contributions from four--fermion
condensates of the form \eqref{C123}. In this paper we have only studied the
effects of the new six--fermion $U(1)$ chiral order parameter. However,
recently, four--fermion operators (which could be associated with the
above--mentioned four--fermion condensates) have been used in the literature,
in the study of \emph{scalar} mesons, which are modelled as four--quark (i.e.,
$\bar{q}q\bar{q}q$) states, called ``\emph{tetraquarks}'' or
``\emph{diquark--antidiquark}'' bound states
\cite{Giacosa2007,tHooft2008,FJS2008-2009}.

\newpage

\renewcommand{\thesection}{}
\renewcommand{\thesubsection}{B.\arabic{subsection} }
 
\pagebreak[3]
\setcounter{section}{1}
\setcounter{equation}{0}
\setcounter{subsection}{0}
\setcounter{footnote}{0}

\begin{flushleft}
{\Large\bf \thesection Appendix B: On the new parameters $F_X$, $\omega_1$
and $c_1$}
\end{flushleft}

\renewcommand{\thesection}{B}

\noindent
The Lagrangian \eqref{newlagr} contains a new field $X$ and three new
parameters, namely $F_X$, $\omega_1$ and $c_1$, with respect to the usual
Lagrangian of Witten, Di Vecchia, Veneziano {\it et al.}
It is therefore natural to ask if the model can be further simplified by
simply eliminating some parameter. As we have already said, in this paper
we are {\it assuming} that the parameter $F_X$, which is essentially
proportional to the new $U(1)$ axial condensate, is different from zero:
in section 3 we discuss the relevance of this parameter $F_X$ in the
phenomenological analysis of the strong decays of pseudoscalar mesons.\\
Concerning the parameter $\omega_1$, we cannot say too much. We remind that
the usual Lagrangian of Witten, Di Vecchia, Veneziano {\it et al.} is
obtained by choosing $\omega_1=1$ (together with $F_X=0$, i.e., $X=0$).
At low temperatures one expects that the deviations from this Lagrangian are
{\it small}, in some sense, and therefore one expects that $\omega_1$ is not
much different from $1$ at low temperatures.
On the other side, as already observed in Ref. \cite{EM1994a}, $\omega_1$ must
necessarily be zero when $T \geq T_{ch}$, in order to avoid a singular
behaviour of the anomalous term above the chiral transition: this implies
a non trivial behaviour of $\omega_1$ with the temperature. However,
in this paper no particular choice for the value of $\omega_1 (T=0)$ will
be done: it will be considered as a {\it free} parameter (apart from the
above--mentioned limitation for $T\geq T_{ch}$).\\
The case of the parameter $c_1$ is much more interesting. By putting $c_1=0$,
i.e., $c=0$ [see Eq. \eqref{c-mu}], into Eqs. \eqref{zl-ql}, these reduce to:
\be
Z_L = \frac{2A[\fp^2 (1-\omega_1)^2 + L \fx^2 \omega_1^2]}{\fp^2 \fx^2} ,~~~
Q_L = 0 ,
\label{zlqlc0}
\ee
which, when inserted into Eq. \eqref{autovalori1}, lead to the following
values for the squared masses of the two singlets $S_1$ and $S_2$
in the chiral limit:
\be
m_{S_1}^2 = 0 ,~~~
m_{S_2}^2 = \left(\frac{2LA}{\fp^2} \right) \omega_1^2 +
\left( \frac{2A}{\fx^2} \right)(1-\omega_1)^2 .
\label{autovaloric0}
\ee
The corresponding eigenvectors, written in terms of $S_\pi$ and $S_X$, are:
\bea
S_1 &=& \frac{1}{\sqrt{\fp^2 (1-\omega_1)^2 + L\fx^2\omega_1^2}}
\left( \fp(\omega_1-1)S_{\pi} + \sqrt{L}\fx \omega_1 S_X \right) , \nonumber \\
S_2 &=& \frac{1}{\sqrt{\fp^2 (1-\omega_1)^2 + L\fx^2\omega_1^2}}
\left( \sqrt{L} \fx \omega_1 S_{\pi} + \fp (1-\omega_1) S_X \right) .
\label{autovect}
\eea
Let us observe that Eqs. \eqref{autovect} and \eqref{autovaloric0} {\it cannot}
be derived by simply putting $c=0$ into Eqs. \eqref{S1-S2} and
\eqref{massaS1-S2}, derived in subsection 2.1.
This is due to the fact that Eqs. \eqref{S1-S2} and \eqref{massaS1-S2} were
derived not only {\it assuming} that $c_1 \neq 0$, but also taking the
large--$N$ limit, in which the quantity $c \equiv \frac{c_1}{\sqrt{2}} \left(
\frac{\fx}{\sqrt{2}}\right) \left(\frac{\fp}{\sqrt{2}}\right)^L$ is
{\it large}, being of order $\OO(N)$ [see Eq. \eqref{relazionin}].
In that case, therefore, one obtains $Z_L = \OO(1)$ and $Q_L = \OO(1/N)$,
so that, from Eq. \eqref{autovalori1}, 
$m_{S_1}^2 \simeq \frac{Q_L}{Z_L} \simeq \frac{2LA}{\fp^2 + L \fx^2} =
\OO(1/N)$ and $S_1$ can be identified with the particle $\eta'$.

Instead, in the particular case in which $c_1=0$ [i.e., $c=0$], one has that
$Z_L = \OO(1/N)$ and $Q_L = 0$, so that, from Eq. \eqref{autovalori1},
$S_1$ is {\it massless} (in the chiral limit) and therefore it does {\it not}
verify the Witten--Veneziano formula required for the $\eta'$.
It is easy to convince oneself that, in this particular case $c_1=0$,
$S_2$, having a squared mass of order $\OO(1/N)$ in the large--$N$ limit,
is just the field which must be identified with the particle $\eta'$,
as required by the Witten--Veneziano mechanism for the solution of the
$U(1)$ problem.
In fact, by virtue of Eqs. \eqref{autovect}, we can re--write the $U(1)$ axial
current $J_{5,\mu}^{(L)}$, given by Eq. \eqref{jmu5}, in terms of the fields
$S_1$ and $S_2$:
\be
J_{5,\mu}^{(L)}=-\sqrt{2L} \partial_{\mu}\left(F_{S_1} S_1 + F_{S_2} S_2
\right),
\label{jmu5c}
\ee
where:
\bea
F_{S_1} &=& \frac{\fp^2 (\omega_1-1) + L \fx^2 \omega_1}
{\sqrt{\fp^2 (1-\omega_1)^2 + L\fx^2 \omega_1^2}} , \nonumber \\
F_{S_2} &=& \frac{\sqrt{L} \fp \fx}
{\sqrt{\fp^2 (1-\omega_1)^2 + L\fx^2 \omega_1^2}} ,
\label{fs1s2}
\eea
are nothing but the {\it decay constants} of the singlet pseudoscalar mesons
$S_1$ and $S_2$, defined as:
\bea
\langle 0 | J_{5,\mu}^{(L)}(0)|S_1 (\vec{p}_1) \rangle &=& i \sqrt{2L} F_{S_1}
p_{1\mu} , \nonumber \\
\langle 0 | J_{5,\mu}^{(L)}(0)|S_2 (\vec{p}_2) \rangle &=& i \sqrt{2L} F_{S_2}
p_{2\mu} .
\label{deffph}
\eea
From Eqs. \eqref{autovaloric0} and \eqref{fs1s2} one immediately
verifies that the field $S_2$ satifies the {\it Witten--Veneziano formula},
i.e.,
\be
m_{S_2}^2= \frac{2LA}{F_{S_2}^2},
\ee
and, therefore, it is nothing but the field associated with the particle
$\eta'$, with a squared (non--chiral) mass generated by the anomaly and
of order $\OO(1/N)$ in the large--$N$ limit, as required by the
Witten--Veneziano mechanism. (Instead, concerning the state $S_1$, even if,
according to Eqs. \eqref{deffph} and \eqref{fs1s2}, it is coupled to the
$U(1)$ axial current, it is {\it not} coupled to the topological charge
density, i.e., $ \langle 0 | Q(0)| S_1(\vec{p}_1) \rangle =
 \frac{1}{\sqrt{2L}} F_{S_1} m^2_{S_1^2}=0 $, since it is massless:
therefore it does {\it not} appear as an intermediate mesonic state in the
spectral decomposition of the {\it full} topological susceptibility $\ldots$)\\
It is interesting to observe that, in this case (differently from the case
discussed in subsection 2.1), the parameter $\omega_1$ plays a fundamental
role. In fact, when $c_1=0$, the {\it anomalous} Lagrangian term containing
$\omega_1$ is the only one which generates a coupling between $U$ and $X$
(i.e., between the {\it usual} quark--antiquark pseudoscalar mesons and the
{\it exotic} singlet state). By changing $\omega_1$ one can ``move''
the anomaly from $U$ to $X$. In particular, in the case $\omega_1=1$ the
anomalous term only depends on $U$ and the field $X$ is decoupled.
In this case the Lagrangian simply reduces to the sum of the usual Lagrangian
written by Witten, Di Vecchia, Veneziano {\it et al.} for the field $U$
(including the anomalous term) {\it plus} a non--anomalous Lagrangian for
the field $X$: in this limit the state $S_2$, i.e, the $\eta'$, reduces to
the usual quark--antiquark singlet state $S_\pi$, while the massless state
$S_1$ reduces to the exotic state $S_X$.
On the contrary, in the opposite case $\omega_1=0$ the anomalous term only
depends on the exotic field $X$ and so the state $S_2$, i.e, the $\eta'$,
reduces to the exotic state $S_X$, while the massless state $S_1$ reduces to
the usual quark--antiquark singlet state $S_\pi$.\\
In conclusion, we have found that, in the case in which $c_1=0$, in addition
to the usual $L^2-1$ non--singlet (pseudo--)Goldstone bosons and to the
{\it massive} singlet $S_2 = \eta'$, there is another singlet $S_1$,
which is {\it massless} in the chiral limit. This particle is therefore
another (pseudo--)Goldstone boson which, when including the quark masses,
should have a mass comparable with that of the other $L^2-1$ non--singlet
pseudoscalar mesons.\\
In the realistic case $L=3$, by diagonalizing the squared mass matrix
\eqref{massmatrix3} with $c=0$, we derive the following expressions for the
squared masses of $\eta$, $S_1$ and $S_2$, at the first order in the
quark masses:
\bea
m^2_\eta &=& \frac{1}{3}B
\biggl\{ (\tilde{m}+2m_s) + (2\tilde{m}+m_s)\alpha_X^2 \nonumber \\
&+& \sqrt{[(\tilde{m}+2m_s) + (2\tilde{m}+m_s)\alpha_X^2]^2
-36\alpha_X^2\tilde{m} m_s} \biggr\} , \nonumber \\
m^2_{S_1} &=& \frac{12\alpha_X^2 B\tilde{m} m_s}
{(\tilde{m}+2m_s) + (2\tilde{m}+m_s)\alpha_X^2 +
\sqrt{[(\tilde{m}+2m_s) + (2\tilde{m}+m_s)\alpha_X^2]^2
-36\alpha_X^2\tilde{m} m_s}} , \nonumber \\ 
m^2_{S_2} &=& \left(\frac{6A}{\fp^2} \right) \omega_1^2 +
\left( \frac{2A}{\fx^2} \right)(1-\omega_1)^2
+ \frac{3 \fx^2 \omega_1^2}{\fp^2(1-\omega_1)^2 + 3\fx^2 \omega_1^2}
\frac{2}{3}B(2\tilde{m}+m_s) ,
\label{align}
\eea
where $\alpha_X^2 \equiv
\frac{\fp^2 (1-\omega_1)^2}{\fp^2 (1-\omega_1)^2 + 3\fx^2 \omega_1^2}$.
Using the fact that $0\le\alpha_X^2\le1$ (and $\tilde{m} < m_s$),
it is easy to show that:
\be
m^2_{S_1} \le 2B\tilde{m} = m^2_\pi \simeq (135 \ \mev)^2 .
\label{eq:aa}
\ee
Even assuming, as already said, that we can identify the singlet $S_2$
with the observed singlet $\eta'$, no other singlet pseudoscalar meson
is observed whose mass satisfies the limit \eqref{eq:aa}. Our assumption
$c_1=0$ (together with $\fx \neq 0$) has thus led us to another
``$U(1)$ problem''.
Even if we let $c_1$ be different from zero, but arbitrarily small, i.e.,
$c_1 \to 0$ with all other quantities fixed, since, by virtue of Eqs.
\eqref{autovalori1}--\eqref{zl-ql}, the squared masses $m^2_{S_1,S_2}$
in the chiral limit are continuous functions of the parameter $c_1$,
we find that $m^2_{S_1} \simeq Q_L/Z_L \simeq
\frac{2Lc}{\fp^2 (1-\omega_1)^2 + L\fx^2\omega_1^2} = \OO(c_1)$
will be arbitrarily small and, when including quark masses, it will have an
upper limit arbitrarily close (from above) to that reported in Eq.
\eqref{eq:aa}.

Therefore, we are forced to discard this possibility (as it leads to wrong
predictions for the pseudoscalar--meson mass spectrum) and, in the rest of
this paper, we shall always consider the model in which $c_1$ is different
from zero and not too small, so that $c=\OO(N)$ is {\it large}.
In this case, as we have seen in subsections 2.1 and 2.2, the squared masses
of the singlet mesons $S_1$ and $S_2$ are given by Eq. \eqref{massaS1-S2}
in the chiral limit and by Eqs. \eqref{mass2} and \eqref{mass3} in the
realistic case with $L=3$ light quark flavours. Therefore, as already said,
the state $S_1$ has a {\it topological} (non--chiral) squared mass of order
$\OO(1/N)$ in the large--$N$ limit and it is nothing but the particle $\eta'$.
Instead, the state $S_2$ is identified with an {\it exotic} singlet particle
$\eta_X$, having a {\it large} (non--chiral) mass term of order $\OO(1)$ in
the large--$N$ limit, generated by the (non--zero) coupling constant $c_1$.

\newpage


\end{document}